\documentclass[preprint,onecolumn,nofootinbib]{revtex4}
\pdfoutput=1
\usepackage{graphicx}%Include figure files
\usepackage{amsmath}
\usepackage{amsfonts}
\usepackage{amssymb,ulem}
\usepackage{color}%
\usepackage{dcolumn}% Align table columns on decimal pointl

\usepackage{MnSymbol,wasysym}
\usepackage{braket}

\usepackage{eurosym}
\usepackage{calrsfs}
\usepackage[usenames,dvipsnames,svgnames]{xcolor}

\newcommand{\RNum}[1]{\uppercase\expandafter{\romannumeral #1\relax}}
\usepackage[colorlinks=true,linkcolor=blue,urlcolor=blue,filecolor=black,citecolor=red,pdfstartview=FitV,pdftitle={},pdfsubject={},pdfkeywords={},pdfpagemode=None,bookmarksopen=true]{hyperref}

\begin{document}
\baselineskip=0.5 cm
\title{Complexity growth of massive black hole with a probe string}
\author{Yu-Ting Zhou$ ^{1,2}$}
\email{constaantine@163.com}
\author{Xiao-Mei Kuang$ ^{2,3}$}
\email{xmeikuang@yzu.edu.cn (corresponding author)}
\author{Jian-Pin Wu$ ^{2,3}$}
\email{jianpinwu@yzu.edu.cn}
\affiliation{$^1$ College of Mathematics and Science, Yangzhou University, Yangzhou 225009, China}
\affiliation{$^2$ Center for Gravitation and Cosmology, College of Physical Science and Technology,
Yangzhou University, Yangzhou 225009, China}
\affiliation{$^3$ School of Aeronautics and Astronautics, Shanghai Jiao Tong University, Shanghai 200240, China}
\date{\today }

\begin{abstract}
\vspace*{0.6cm}
\baselineskip=0.5 cm
In this work, we study the computational complexity of massive gravity theory via the ``Complexity = Action" conjecture. Our system contains a particle moving on the boundary of the black hole spacetime. It is dual to inserting a fundamental string in the bulk background. Then this string would contribute a Nambu-Goto term, such that the total action is composed of the Einstein-Hilbert term, Nambu-Goto term and the boundary term. We shall investigate the time development of this system, and mainly discuss the features of the Nambu-Goto term affected by the graviton mass and the horizon curvature in different dimensions. Our study could contribute interesting properties of complexity.
\end{abstract}

\maketitle
\tableofcontents

\section{Introduction}\label{sec:Introduction}
 Along with the development of fundamental physics and information theory, some very remarkable tools were proposed to reveal the mystery of nature. One of them is in the areas of general relativity and quantum field theory. In 1990s, the AdS/CFT duality which connects the gravity theory and its boundary conformal field theory was proposed \cite{Maldacena:1997re,Gubser:1998bc, Witten:1998qj} and it has opened a new avenue for us to investigate the gravity in the framework of holography.

Thanks to holography, the entanglement entropy (EE), which measures the degrees of freedom in a strongly coupled system originally, has a holographic description, stating that the EE for a subregion on the dual boundary is proportional to the minimal Hubeny-Rangamani-Takayanagi surface in the bulk geometry\cite{Takayanagi:2012kg,Hubeny:2007xt}.
%
%Later, Susskind et al pointed out in \cite{Stanford:2014jda} that the Einstein-Podolsky-Rosen(EPR) correlation can be views as a Einstein-Rosen(ER) bridge (worm hole), which called the "ER=EPR" conjecture. This conjecture states that  people from two sides of wormhole can exchange information each other.
%
The other significant concept in theoretical physics is the computational complexity which originally comes from the quantum information theory\cite{Osborne:2012hc,Gharibian:2015qhc,Dvali:2017ubs,Swingle:2016var,Hashimoto:2017fga,Watrous:2008qc,Bao:2018ira}. The computational complexity measures the difficulty of turning a quantum state into another state. However, from the side of the field theory, it is extremely difficult to evaluate it when the degrees of freedom of the system becomes large. It is also not clear on how to precisely define the initial state and reference state. Though there are plenty of works on this theme from quantum field theory \cite{Vanchurin:2016met,Chapman:2017rqy,Jiang:2018gft,Molina-Vilaplana:2018sfn,Bhattacharyya:2018wym} as well as works on geometric way to define the complexity \cite{Nielsen:2005ap,Nielsen:2006qcg,Nielsen:2007tgqc,Jefferson:2017sdb,Yang:2018nda,Khan:2018rzm,Hackl:2018ptj}, a well-defined theory of complexity is still unknown.

On the other hand, in gravity side, only considering the outside properties of black hole is not enough for us to understand the interior black hole. Susskind \emph{et al} suggested  that the computational complexity can measure the size of wormhole which anchored at the two sides of boundary times of AdS black hole\cite{Stanford:2014jda,Susskind:2014moa,Susskind:2014jwa,Susskind:2014rva,Susskind:2018pmk}. Accordingly, there are two conjectures have emerged. One is ``Complexity=Volume"(CV) conjecture. $V$ denotes the volume of Einstein-Rosen(ER) bridge that connects the two side of boundary times of AdS black hole. The other reliable candidate proposal is the ``Complexity=Action"(CA) conjecture. $A$ denotes the classical action of a spacetime region which was enclosed by the bulk Cauchy slice that anchored at the boundaries, and this domain was also called ``Wheeler-Dewitt patch"\cite{Brown:2015bva,Brown:2015lvg}.

Especially, the application of CA conjecture have been extensively studied in \cite{Pan:2016ecg,Momeni:2016ekm,Chapman:2016hwi,Lehner:2016vdi,Carmi:2016wjl,Tao:2017fsy,Alishahiha:2017hwg,
Reynolds:2017lwq,Qaemmaqami:2017lzs,Couch:2017yil,Guo:2017rul,Sebastiani:2017rxr,Swingle:2017zcd,Cano:2018aqi,
Chapman:2018dem,Chapman:2018lsv,Auzzi:2018pbc,Ghaffarnejad:2018prc,Alishahiha:2018tep,An:2018xhv,Cai:2016xho,Babaei-Aghbolagh:2020vsz} and therein. However, the aforementioned work focused on the stationary system. An important direction is to generalize the CA conjecture
into a time-dependent process such as a system moved by a drag force, which  was motivated by the work of jet-quenching phenomena in ion collisions where the system is moved by a drag force\cite{Gubser:2006bz}. When a charged particle passes through a quark-gluon plasma, it would lose energy because of the effect of shear viscosity. Thus, the most useful method to analyze this dissipative system is to use a probe. Many important work were inspired by this clue. For instance, in \cite{Abad:2017cgl} the growth of complexity with the probe brane was studied. In \cite{Ageev:2014nva} the  non-local operator was studied in BTZ black hole. More recently, Nagasaki investigated the complexity of  AdS$_5$ black hole via CA by inserting a probe string which moves on a circle in the spacial part of the AdS spacetime\cite{Nagasaki:2017kqe}. This study was also extended in rotating black holes\cite{Nagasaki:2018csh,Nagasaki:2019icm} and the black holes in Horndeski gravity\cite{Santos:2020xox,Santos:2020lmb}.

In this work, we also use this method and investigate the complexity growth in massive gravity with a probe string.
Massive gravity theory is a theory beyond Einstein theory of gravity where the graviton is massless. Recently, significant progress has been made towards constructing massive gravity theories that avoid instability, see for example\cite{deRham:2010ik,deRham:2010kj,Hassan:2011hr,Hassan:2011vm,Hassan:2011tf,Desai:2010ea,Ludeling:2012cu}. More importantly, it was addressed in \cite{Blake:2013bqa} that the massive terms in the gravitational action break the diffeomorphism symmetry in the bulk, which corresponds to momentum dissipation in the dual boundary field theory.
Moreover, the complexity growth for the stationary massive black hole has been studied in \cite{Pan:2016ecg,Guo:2017rul}, in which the complexity growth stems from the contribution from Einstein-Hilbert action and the boundary term.

The aim of this work is to study the effect of probe string on the complexity growth in massive gravity.  This means that we shall consider a Wilson line operator in the theory, which is dual to a dynamical system. The novelty of this work is that the non-local operator could correspond to a particle moving on the boundary gauge theory with momentum relaxation. We shall focus on this effect of Wilson line operator on the Nambu-Goto action. In details, we shall study the influence of horizon curvatures, the graviton mass and the dimension of spacetime on the velocity dependent complexity growth, which is dual to Nambu-Goto action growth in massive black hole with a probe string.

This paper is organized as follows. In section \ref{2} we  briefly review the massive gravity and then derive the Nambu-Goto action in massive gravity with probe string. In section \ref{3} and section \ref{4}, we investigate the Nambu-Goto action growth in the three and higher dimensional gravity, respectively. We analyze the effect of horizon curvature and graviton mass on the features of the Nambu-Goto action growth. We summarize in section \ref{summ}. We shall use natural units with $G = \hbar = c = 1$.
%%%%%%%%%%%
%%%%%%%%%%%
\section{Massive gravity and the Nambu-Goto action} \label{2}
\subsection{Review of massive black hole}
We consider $(n+2)$-dimensional action of massive gravity\cite{Vegh:2013sk,Cai:2014znn},
\begin{equation}
S=\frac{1}{16\pi G} \int d^{n+2} x \sqrt{-g}\left[R+\frac{n(n+1)}{l^{2}}+m^{2} \sum_{i}^{4} c_{i} \mathcal{U}_{i}(g, f)\right] ,
\label{MssGraAction}
\end{equation}
where $R$ is the scalar curvature, $l$ is the AdS radius and $f$ is a fixed symmetric tensor. In addition, $c_{i}$ is constant and $\mathcal{U}_{i}$ is the symmetric polynomial of the
eigenvalue of the $(n+2)\times (n+2) $ matrix $\mathcal{K}_{\nu }^{\mu }=\sqrt{%
g^{\mu \alpha }f_{\alpha \nu }}$ which can be written as follows
\begin{eqnarray}
\mathcal{U}_{1} &=&\left[ \mathcal{K}\right] ,\;\;\;\;\;\mathcal{U}_{2}=%
\left[ \mathcal{K}\right] ^{2}-\left[ \mathcal{K}^{2}\right] ,\;\;\;\;\;%
\mathcal{U}_{3}=\left[ \mathcal{K}\right] ^{3}-3\left[ \mathcal{K}\right] %
\left[ \mathcal{K}^{2}\right] +2\left[ \mathcal{K}^{3}\right] ,  \notag \\
&&\mathcal{U}_{4}=\left[ \mathcal{K}\right] ^{4}-6\left[ \mathcal{K}^{2}%
\right] \left[ \mathcal{K}\right] ^{2}+8\left[ \mathcal{K}^{3}\right] \left[
\mathcal{K}\right] +3\left[ \mathcal{K}^{2}\right] ^{2}-6\left[ \mathcal{K}%
^{4}\right] ,
\end{eqnarray}
where $[\mathcal{K}]=\mathcal{K}_{\mu}^{\mu}$ and the square root in $\mathcal{K}$ can be interpreted as $(\sqrt{\mathcal{K}})^{\mu}_{~\nu}(\sqrt{\mathcal{K}})^{\nu}_{~\lambda}
=\mathcal{K}^{\mu}_{~\lambda}$.
As known that the presence of impurities in realistic materials leads that the momentum is not conserved, so that the system gives finite DC conductivity. Modeling systems via translationally invariant quantum field
theories always comes across problems unless the effects of momentum dissipation is incorporated.
In holographic framework, several models have been proposed to involve momentum dissipation, which brings in finite DC conductivity.
Massive gravity is a completive candidate in which the momentum dissipation is involved, and it is an effective bulk theory that does
not conserve momentum without borrowing additional fields.
In the action (\ref{MssGraAction}), the last terms represent massive potentials associated with the graviton mass which breaks the diffeomorphism invariance in the bulk, which produces momentum relaxation in the dual boundary theory.

Variating the action, we can obtain the equations of motion
\begin{equation}
\begin{aligned} R_{\mu \nu}-\frac{1}{2} R g_{\mu \nu}-\frac{n(n+1)}{2 l^{2}} g_{\mu \nu}-\frac{1}{2}\left(F_{\mu \sigma} F_{\nu}^{\sigma}-\frac{1}{4} g_{\mu \nu} F^{2}\right)+m^{2} \chi_{\mu \nu} &=0 \end{aligned},
\label{Field equation}
\end{equation}%
where
\begin{eqnarray}
\begin{aligned} \chi_{\mu \nu}=&-\frac{c_{1}}{2}\left(\mathcal{U}_{1} g_{\mu \nu}-\mathcal{K}_{\mu \nu}\right)-\frac{c_{2}}{2}\left(\mathcal{U}_{2} g_{\mu \nu}-2 \mathcal{U}_{1} \mathcal{K}_{\mu \nu}+2 \mathcal{K}_{\mu \nu}^{2}\right)-\frac{c_{3}}{2}\left(\mathcal{U}_{3} g_{\mu \nu}-3 \mathcal{U}_{2} \mathcal{K}_{\mu \nu}\right.\\ &\left.+6 \mathcal{U}_{1} \mathcal{K}_{\mu \nu}^{2}-6 \mathcal{K}_{\mu \nu}^{3}\right)-\frac{c_{4}}{2}\left(\mathcal{U}_{4} g_{\mu \nu}-4 \mathcal{U}_{3} \mathcal{K}_{\mu \nu}+12 \mathcal{U}_{2} \mathcal{K}_{\mu \nu}^{2}-24 \mathcal{U}_{1} \mathcal{K}_{\mu \nu}^{3}+24 \mathcal{K}_{\mu \nu}^{4}\right) \end{aligned} \label{Xmunu}
\end{eqnarray}

The static black hole solution of the above action yields
\begin{equation}
d s^{2}=-f(r) d t^{2}+f^{-1}(r) d r^{2}+r^{2} h_{i j} d x^{i} d x^{j}, \quad i, j=1,2,3, \cdots, n,  \label{metric}
\end{equation}
where $h_{i j} d x^{i} d x^{j}$ is a line element of Einstein space with constant curvature $n(n-1)k$ and $k=1,0,-1$  corresponds to a spherical, Ricci flat, and hyperbolic horizon  for black hole. We  follow the ansatz in \cite{Cai:2014znn} and set the reference metric $f_{\mu \nu }=diag(0,0,c_{0}^{2}h_{ij})$. Then
 $\mathcal{U}_{i}$ can be computed as
\begin{equation}
\begin{aligned} \mathcal{U}_{1} &=n c_{0} / r,~~~ \mathcal{U}_{2} =n(n-1) c_{0}^{2} / r^{2}, \\
\mathcal{U}_{3} &=n(n-1)(n-2) c_{0}^{3} / r^{3},~~~ \mathcal{U}_{4} =n(n-1)(n-2)(n-3) c_{0}^{4} / r^{4}. \end{aligned}  \label{U}
\end{equation}
Putting them back to the Einstein equation, we have the metric function $f(r)$
\begin{equation}
\begin{aligned}f(r)=& k+\frac{r^{2}}{l^{2}}-\frac{m_{0}}{r^{n-1}}+\frac{c_{0}c_{1}m^2}{n} r  +c_{0}^{2}c_{2}m^{2}+\frac{(n-1)c_{0}^{3}c_{3}m^{2}}{r}+\frac{(n-1)(n-2)c_{0}^{4}c_{4}m^{2}}{r^{2}}, \end{aligned} \label{fr}
\end{equation}
where the integral constant $m_0$ is the black hole mass parameter.

\subsection{Nambu-Goto action}
Following the approach in \cite{Nagasaki:2017kqe,Nagasaki:2018csh}, we consider a Wilson line
operator in the spacetime by inserting a fundamental string in the massive gravity. This corresponds to a test particle moving on the boundary gauge theory, which is described by a non-local operator. Inserting the Wilson loop is described by adding a Nambu-Goto (NG) term, so the total action consists of the  Einstein-Hilbert term, the Nambu-Goto term and the boundary term.
Since the contribution of  Einstein-Hilbert action and boundary term to the complexity growth in massive black hole has been studied in \cite{Pan:2016ecg,Guo:2017rul}, here we shall investigate the effect of the Nambu-Goto action.

\underline{\textbf{Setup}} To this end, we assume that a probe string moves in a  subspace with different topologies. Then the induced metric was $h_{i j} d x^{i} d x^{j}=d\phi^{2}$. We take the worldsheet parameter as
\begin{equation}
t=\tau \quad \text{,} \quad r=\sigma \quad \text{,} \quad \phi=v\tau+\xi(\sigma). \label{worldsheetpara}
\end{equation}
where $v$ denotes the velocity of a string relative to the black hole. For simplicity,  we will set $l$=1, then the induced metric is
\begin{equation}
\begin{aligned}&ds_{n+2(ind)}^{2}=(-f(\sigma)+\sigma^{2}v^{2})d\tau^{2}+(\frac{1}{f(\sigma)}+\sigma^{2}\xi^{\prime}(\sigma))d\sigma^{2}+2\sigma^{2}v\xi^{\prime}d\tau d\sigma, \\& f(\sigma)=k+\sigma^{2}-\frac{m_{0}}{\sigma^{n-1}}+\frac{c_{0}c_{1}m^{2}}{n} \sigma  +c_{0}^{2}c_{2}m^{2}+\frac{(n-1)c_{0}^{3}c_{3}m^{2}}{\sigma}+\frac{(n-1)(n-2)c_{0}^{4}c_{4}m^{2}}{\sigma^{2}}. \end{aligned} \label{n+2metric2}
\end{equation}
The NG action is achieved by integrating over the WDW patch,
\begin{equation}
\begin{aligned}\frac{dS_{NG}}{dt}=&T_{s}\int_{0}^{r_{h}}d\sigma \sqrt{-g_{ind}(\sigma)}=T_{s}\int_{0}^{r_{h}}d\sigma \sqrt{1-\frac{v^{2}\sigma^{2}}{f(\sigma)}+\sigma^{2}f(\sigma)\xi^{\prime}(\sigma)^{2}}\equiv\int_{0}^{r_{h}}d\sigma \mathcal{L}_{n+2}\end{aligned} \label{NGaction}
\end{equation}
where the horizon $r_{h}$ is determined by $f(r_h)=0$, and $T_s$ is the tension of fundamental string.

Varying the above `action', we obtain the equation of motion for $\xi$
\begin{equation}
\frac{d}{d\sigma}\left(\frac{\sigma^{2}f(\sigma)\xi^{\prime}(\sigma)}{\sqrt{1-v^{2}\sigma^{2}/f(\sigma)
+\sigma^{2}f(\sigma)\xi^{\prime}(\sigma)^{2}}}\right)=0, \label{eomxi}
\end{equation}
from which we could define the constant $c_{\xi}$ as
\begin{equation}
c_{\xi}\equiv\frac{\sigma^{2}f(\sigma)\xi^{\prime}(\sigma)}{\sqrt{1-v^{2}\sigma^{2}/f(\sigma)
+\sigma^{2}f(\sigma)\xi^{\prime}(\sigma)^{2}}}. \label{cxi}
\end{equation}
Subsequently, $\xi^{\prime}$ can be solved by
\begin{equation}\label{xiprime}
\xi^{\prime}(\sigma)=\frac{c_{\xi}}{\sigma f(\sigma)}\sqrt{\frac{f(\sigma)-v^{2}\sigma^{2}}{\sigma^{2}f(\sigma)-c_{\xi}^{2}}}.
\end{equation}
This expression must give real values, so the denominator and numerator in the square root should coincidentally  be
positive,negative or zero. It is noted that we shall set $c_0=c_1=c_2=c_3=c_4=1$ without loss of generality. {Then regarding the numerator, $f(\sigma)-v^{2}\sigma^{2}$, as a function of $\sigma$, we find it is a monotonically increasing function and has a negative value for $\sigma=0$. Therefore, there is only one solution $\sigma=\sigma_{H}$ to  $f(\sigma)-v^{2}\sigma^{2}=0$.
Subsequently, the constant $c_{\xi}$ is determined by $\sigma_{H}^{2}f(\sigma_{H})-c_{\xi}^{2}=0$ as}
\begin{equation}\label{cxi}
c_{\xi}=v\sigma_{H}^{2}.
\end{equation}
Inserting the expressions \eqref{n+2metric2},\eqref{xiprime} and \eqref{cxi} into \eqref{NGaction},
we could rewrite the NG action as
\begin{eqnarray}\label{NG2}
&&\frac{1}{T_{s}}\frac{dS_{NG}}{dt}=\int_{0}^{r_{h}}d\sigma \sqrt{\frac{g(\sigma)-\sigma^{4}v^{2}}{g(\sigma)-\sigma_{H}^{4}v^{2}}}~~\mathrm{with}~~ \\
&& g(\sigma)=k\sigma^{2}+\sigma^{4}-m_{0}\sigma^{3-n}+\frac{c_{0}c_{1}m^{2}}{n}\sigma^{3}
+c_{0}^{2}c_{2}m^{2}\sigma^{2}+(n-1)c_{0}^{3}c_{3}m^{2}\sigma +(n-1)(n-2)c_{0}^{4}c_{4}m^{2}.\nonumber
\end{eqnarray}
In the following we shall study the effect of horizon curvature and the graviton mass on the NG action in different dimensions.

\section{Three dimensional case } \label{3}
For $n=1$, i.e., the 3-dimensional massive gravity, the terms related with $c_{2},c_{3},$ and $c_{4}$ in \eqref{NG2} vanish, and the NG action would reduce to
\begin{equation}
\begin{aligned}
\frac{1}{T_{s}}\frac{dS_{NG}}{dt}= & \int_{0}^{r_{h}}d\sigma \sqrt{\frac{\sigma^{4}-\sigma^{2}m_{0}+c_{0}c_{1}m^{2}\sigma^{3}-v^{2}\sigma^{4}}{\sigma^{4}
-\sigma^{2}m_{0}+c_{0}c_{1}m^{2}\sigma^{3}-v^{2}\sigma_{H}^{4}}},
\end{aligned}   \label{n3NG}
\end{equation}
where $r_{h}=\frac{-c_{0}c_{1}m^{2}+\sqrt{c_{0}^{2}c_{1}^{2}m^{4}+4m_{0}}}{2}$ and $\sigma_{H}=\frac{-c_{0}c_{1}m^{2}+\sqrt{c_{0}^{2}c_{1}^{2}m^{4}+4(1-v^{2})m_{0}}}{2(1-v^{2})}$.
This integral for all parameter ranges can be directly worked out by numeric. We first turn off the graviton mass and
consider the static BTZ black hole, and then we move onto the massive BTZ case. The numerical results and some analytical study are shown as follows.

\subsection{BTZ black hole}
\begin{figure}[ht!]
 \centering
  \includegraphics[width=5.8cm]{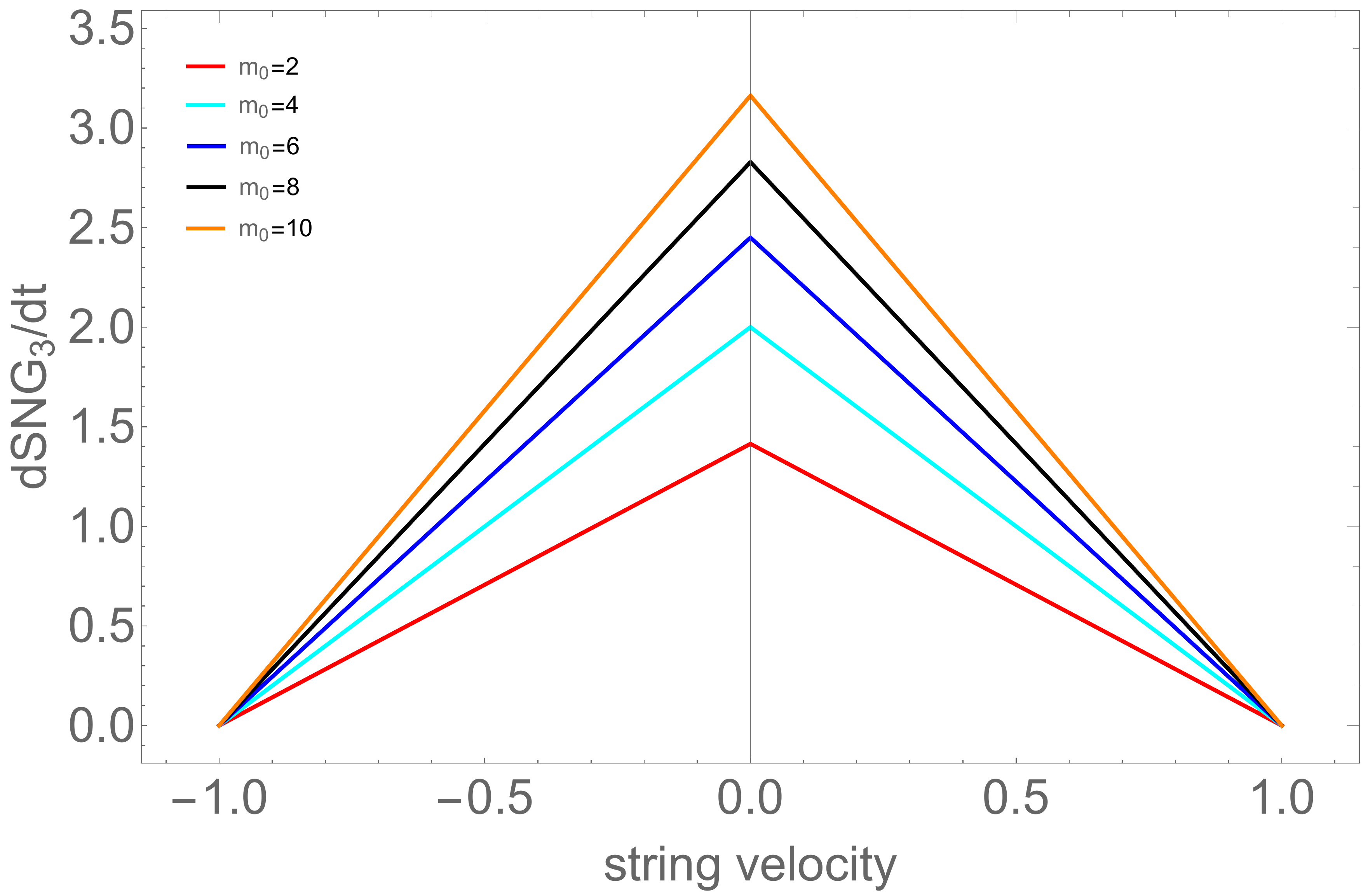}\ \hspace{0.5cm}
  \includegraphics[width=6cm]{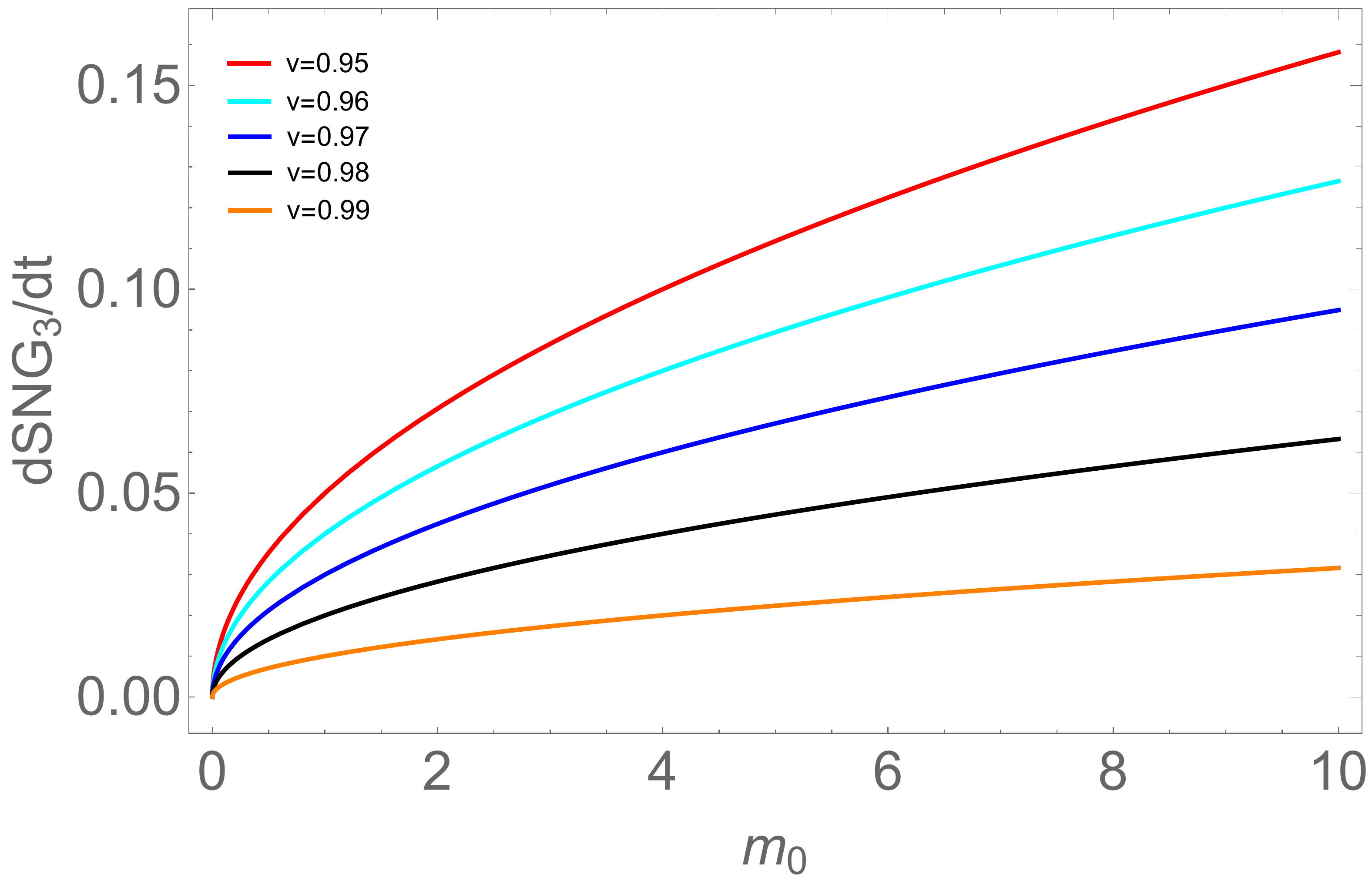}
	  \caption{ Left:{Action growth v.s. String velocity} in BTZ black hole for different $m_{0}$.
	  Right: {Action growth v.s. Black hole mass} in BTZ black hole for different string velocities.}
 \label{fig:d3Mv}
\end{figure}

First we set the graviton mass as $m=0$ and study the effect of BTZ black hole mass on the NG action.
The explicit behaviors with different parameters are plotted in Fig.\ref{fig:d3Mv}. In the left plot  we show the relation between action growth and string velocity with different black hole masses.  We see that the  complexity growth takes the maximum value when the  string is stationary, and as the string moves faster, the action growth becomes smaller. Moreover, the peak value of action growth is larger for larger black hole mass. These features are similar as the observe for static black holes \cite{Nagasaki:2017kqe}and \cite{Nagasaki:2018csh}. Additionally, the action growth will vanish as the velocity approaches to light speed, which was also observed in the rotating BTZ black holes \cite{Nagasaki:2018csh} even though the peak is shifted.
%%%%%%

%%%%%%
The right plot of Fig.\ref{fig:d3Mv} shows the relation between action growth and black hole mass with different string velocities. In this case, the faster string gives the smaller action growth, which is consistent with that in the left plot. It behaves as a monotonically increasing function of the black hole mass, which is similar to the observes in rotating BTZ black hole \cite{Nagasaki:2017kqe}. A possible interpretation is that a lager system may have larger information and so it is more complex.

The above numerical results can be analytically confirmed. With tricks,  the growth of NG action \eqref{n3NG} could be analytically integrated as
\begin{equation}
\frac{1}{T_{s}}\frac{d S_{NG}}{dt}=\sqrt{m_{0}}-\lvert v\rvert \sqrt{m_{0}},
\end{equation}
which depends on the black hole mass and string velocity in a simple way. It indicates that with fixed $m_0$, the action growth is maximum when the string is stationary with $v=0$ and the maximal value is $\sqrt{m_0}$. As $\lvert v\rvert$ increases, it decreases and could vanish when the speed approaches to the light speed with $\lvert v\rvert=1$.  Moreover, the action growth could increase monotonously as $m_{0}$ when $v$ is fixed.

\subsection{Massive BTZ black hole}
\begin{figure}[ht!]
 \centering
  \includegraphics[width=5.8cm]{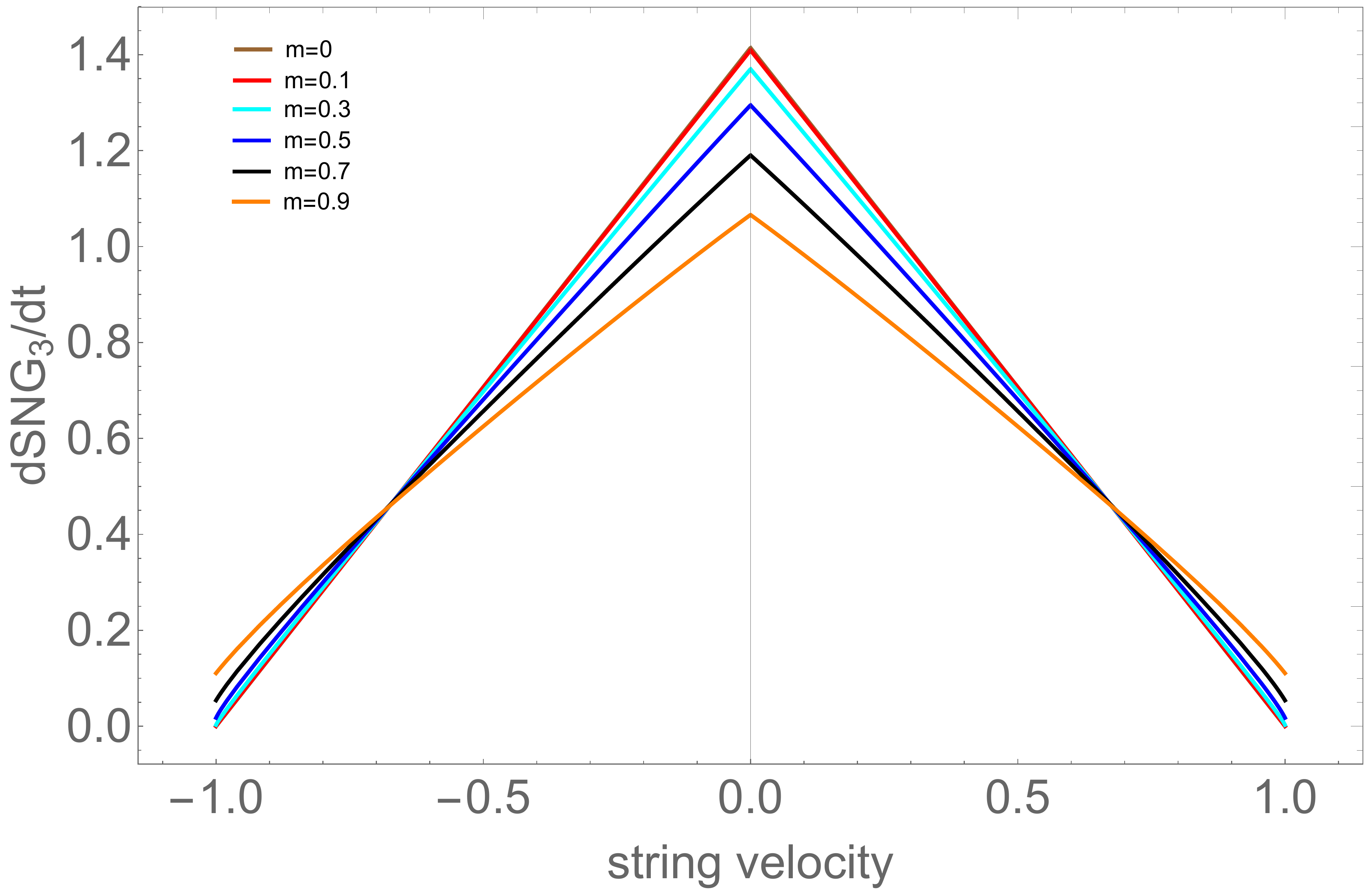}\ \hspace{0.5cm}
  \includegraphics[width=6cm]{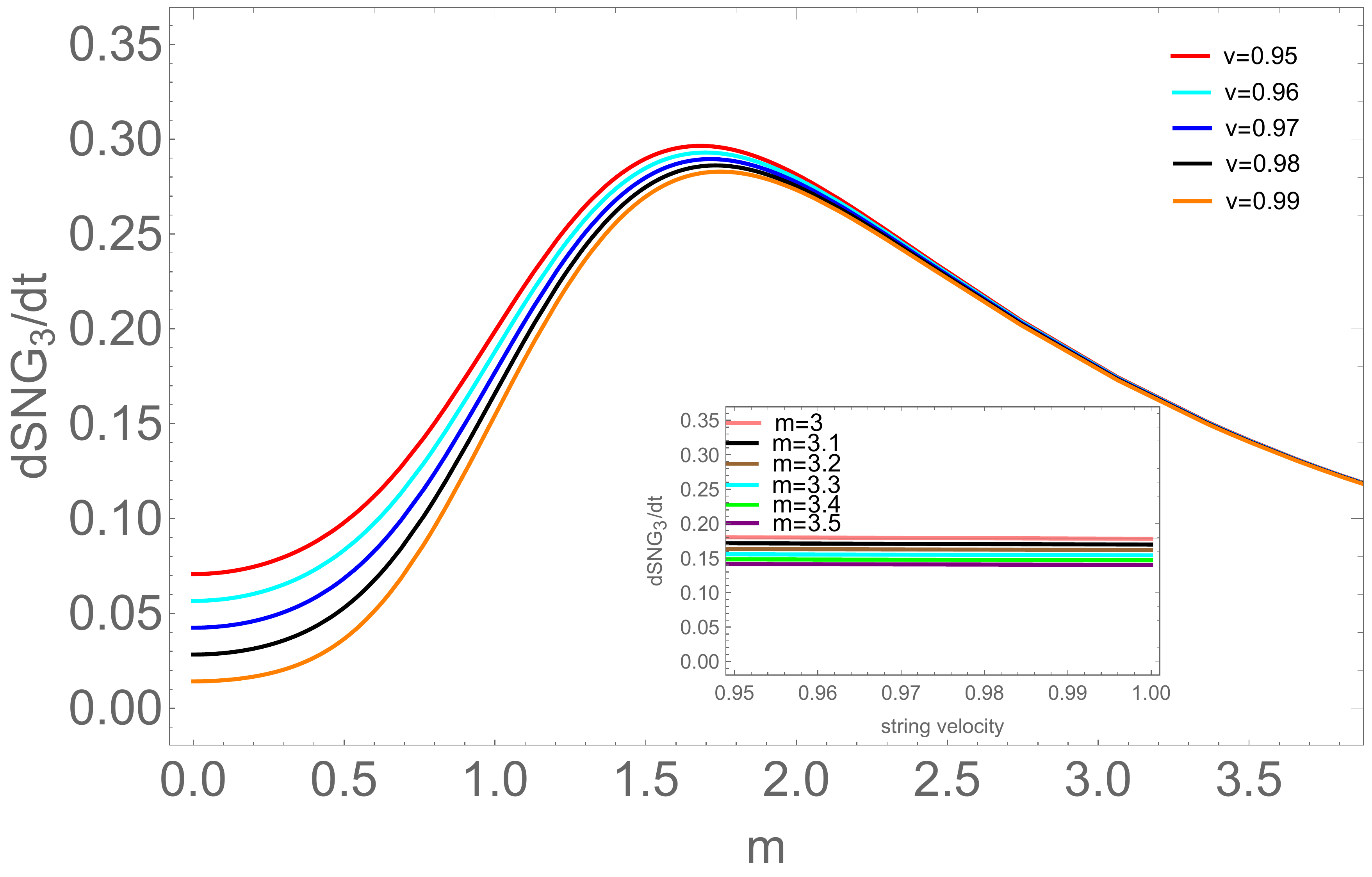}\ \hspace{0.5cm}
	  \caption{Left:  {Action growth v.s. String velocity} in massive BTZ black hole.  Right: {Action growth v.s. Graviton mass}  in massive BTZ black hole. Here we fix the black hole mass $m_{0}=2$.}
 \label{fig:d3k0}
\end{figure}

Then we turn on the graviton mass and  study its effect on the NG action growth. So we fix the black hole mass as $m_0=2$. The results are shown in Fig.\ref{fig:d3k0}.
%%%%%%%
%%%%%%%%

In left plot of Fig.\ref{fig:d3k0}, we show the relation between action growth and string velocity for  massive black hole. As usual, the maximal complexity growth appears when  the string velocity is equal to zero for different $m$. The interesting point is that the maximal value in massive BTZ black hole is smaller than that in BTZ black hole, and as the graviton mass increases, the maximal value becomes smaller. For all chosen $m$, as the string moves faster, the action growth becomes smaller. Moreover, comparing to the BTZ case, the action growth will not vanish when the string moves with light speed. In addition, It is also obvious that the effect of $m$ on the action growth depends on the velocity.

Then in the right plot of  Fig.\ref{fig:d3k0}, we present the relation between action growth and the graviton mass. We focus on the vicinity of the light speed. In all cases, as $m$ increases, the action growth increases into a peak and then decreases, i.e., it is not a monotonically increasing function. Moreover, the velocity dependence is significant for small graviton mass, but it seems to be slight for large $m$, which is also explicit in the inserted plot. Thus, on the vicinity of the light speed, the graviton mass suppresses the velocity dependence of the action growth, which is very different from the effect of black hole mass that enhances it in all cases as studied in \cite{Nagasaki:2017kqe,Nagasaki:2018csh}.
%%%%%
\begin{figure}[ht!]
 \centering
  \includegraphics[width=7cm]{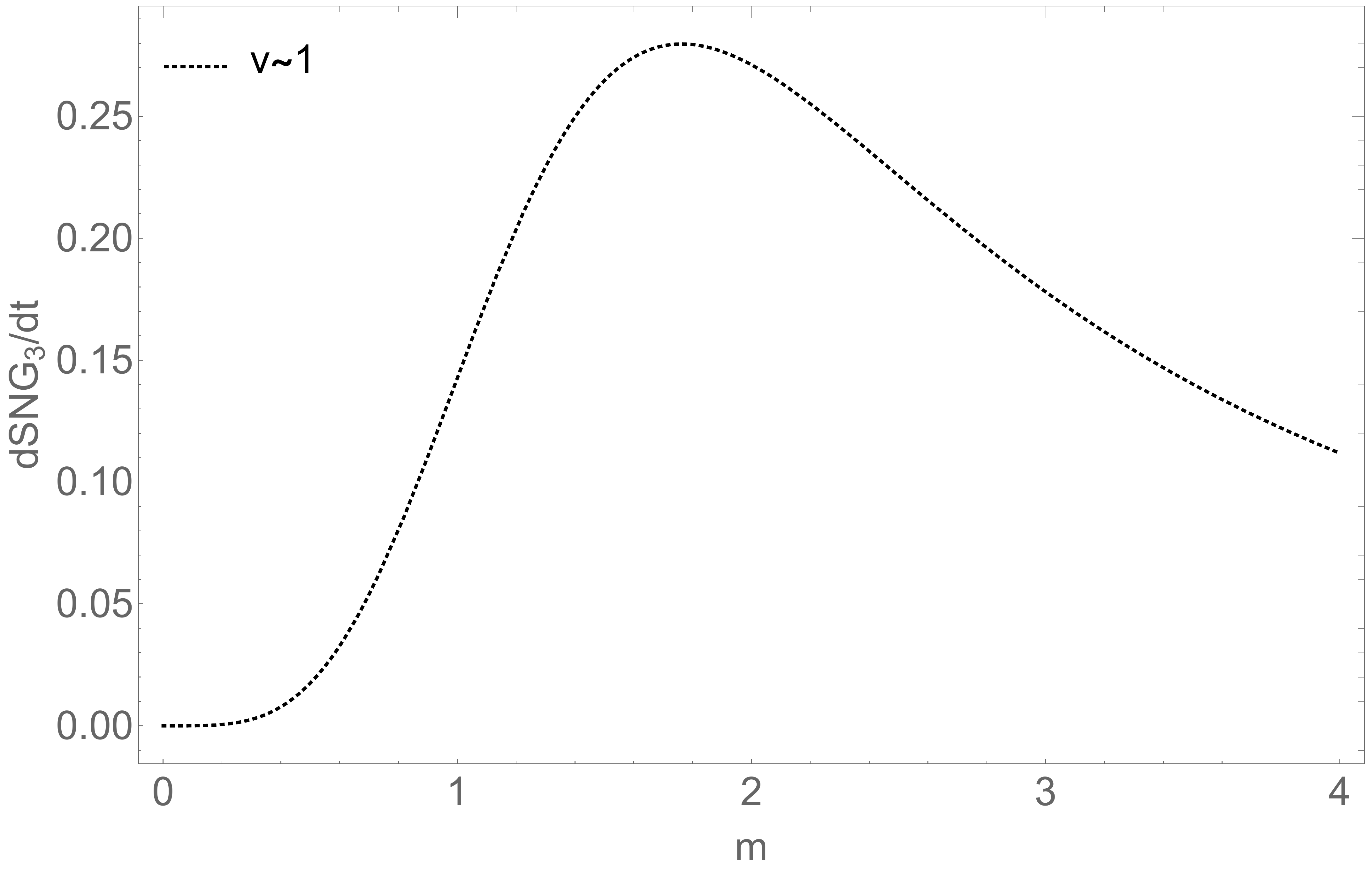}
  \caption{Action growth v.s. Graviton mass in massive BTZ black hole. Here we fix the black hole mass $m_{0} = 2$ and $v\sim 1 $.}
 \label{fig:d3semianam}
\end{figure}

It is noted that in this case, it is difficult to integrate the action \eqref{n3NG} analytically. However, we could  estimate the behavior in some limit regimes to further check the numerical analysis.
In the limit $v \rightarrow 0$, we can integrate \eqref{n3NG} as
\begin{eqnarray}\label{3danalyz1}
\left .\frac{1}{T_{s}}\frac{dS_{NG}}{dt} \right|_{v=0} = \frac{\sqrt{m^{4}+4m_{0}}-m^{2}}{2},
\end{eqnarray}
which is a monotone decreasing function of positive $m$. So the value of the action growth at $v=0$ is smaller as the graviton mass increases, which is consistent with that shown in the left plot of Fig.\ref{fig:d3k0}.

In the speed light limit with $v \rightarrow 1$, a semi-analytic integration is obtained as
\begin{eqnarray}\label{3danalyz2}
\left .\frac{1}{T_{s}}\frac{dS_{NG}}{dt} \right|_{v=1} \sim m\int_{0}^{r_{h}}\frac{\sigma d\sigma}{\sqrt{(\sigma+\frac{m_{0}}{m^{2}})(\sigma^{2}+(\frac{m_{0}}{m^{2}})^{2})+m^{2}\sigma^{2}}},
\end{eqnarray}
which is shown in Fig.\ref{fig:d3semianam}. It shows that in this limit, as $m$ increases, the action growth increases into a peak and then decreases, i.e., it is not a monotonic function, which matches the numerical study shown in the right plot of Fig.\ref{fig:d3k0}.

%%%%%%%%%%%%
%%%%%%%%%%%%
\section{Higher dimensional case} \label{4}
When the dimension is higher than three, the black hole horizon could have hyperbolic, plane and spherical topologies corresponding to  the horizon curvatures $k=-1,0$ and $1$, respectively. In higher dimensions, the exact integration results of \eqref{NG2} is difficult, so we shall numerically integrate it in different dimensions and study the effect the horizon curvature  and graviton mass on the NG action growth.
%%%%%%%%%%%%
%%%%%%%%%%%%
\subsection{AdS black hole }
We first study the effect of horizon curvature. So we focus on the AdS black hole by setting the graviton mass as $m=0$.
%%%%%%%%%%%%%%%%%%%%%%%%
\begin{figure}[ht!]
 \centering
  \includegraphics[width=5.8cm]{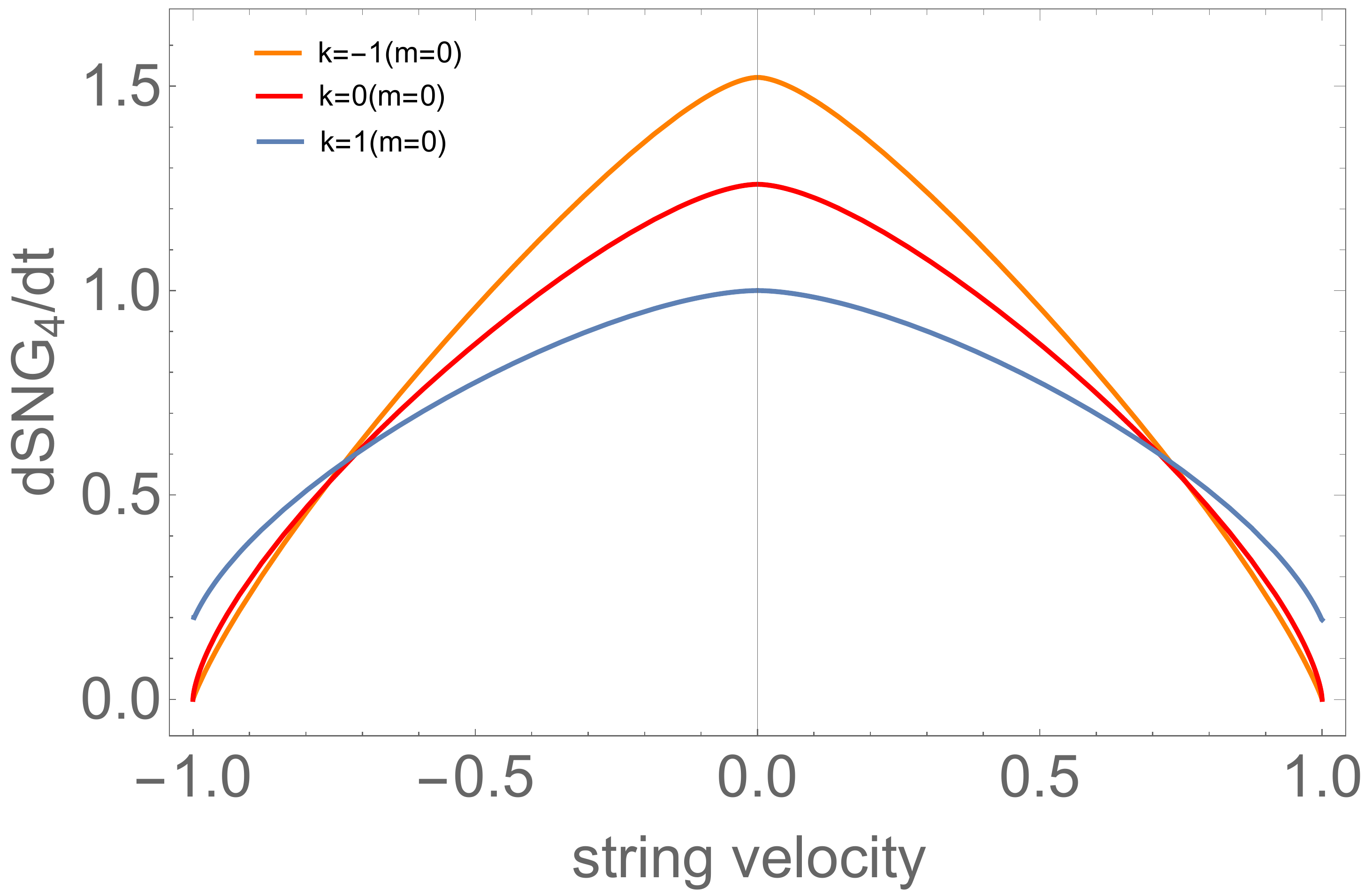}\hspace{1cm}
  \includegraphics[width=5.9cm]{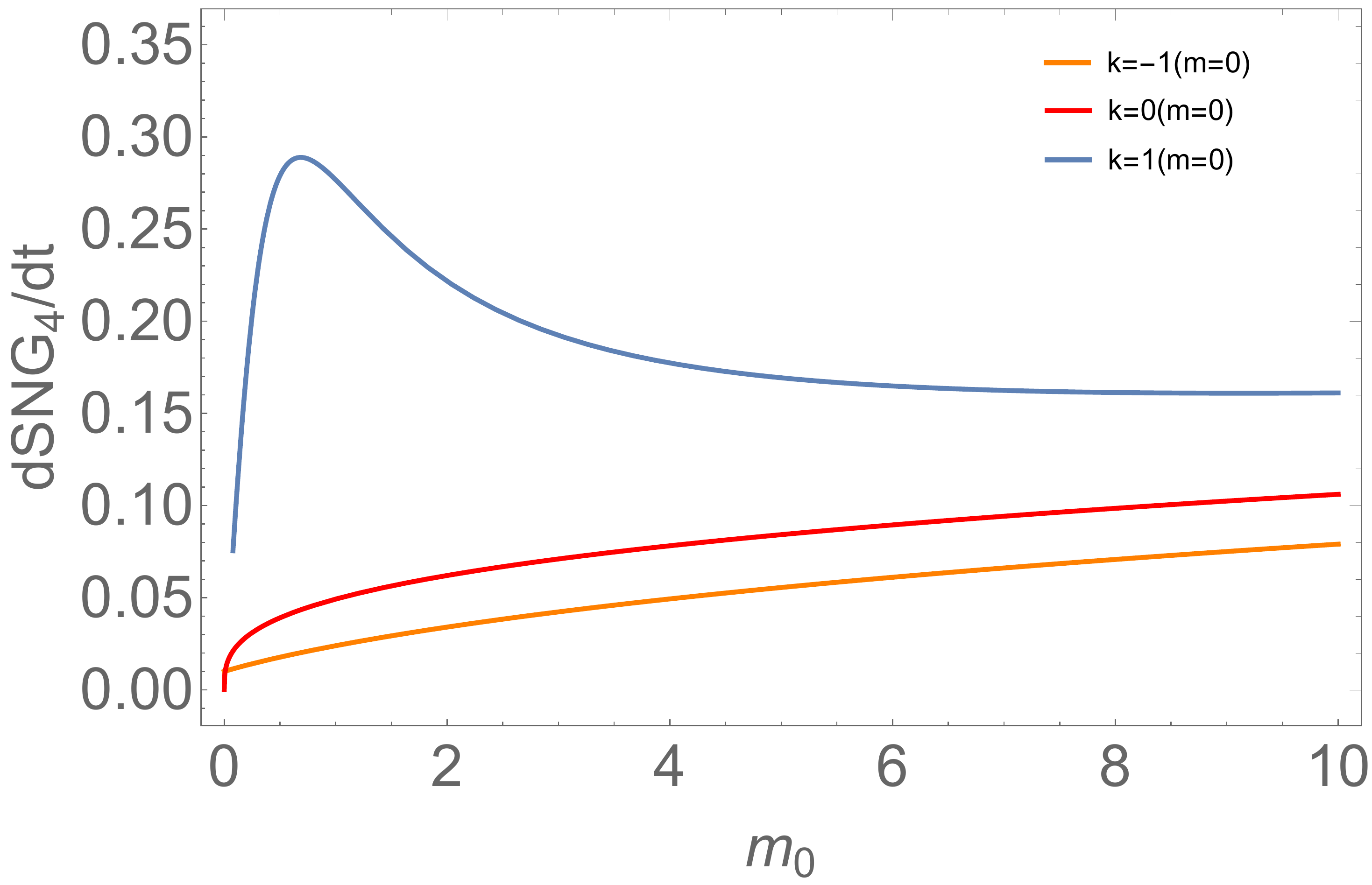}
	  \caption{Left: {Action growth v.s. String velocity } for AdS$_4$ black hole. Here we have fixed  $m_{0}=2$ and $m=0$.
Right: {Action growth v.s. Black hole mass } for AdS$_4$ black hole. Here we fixed $v=0.99$.}
 \label{fig:d4m0kv}
\end{figure}
%%%%%%%%%%%%
%%%%%%%%%%%%
%%%%%%%
%%%%%%%
\begin{figure}[ht!]
 \centering
  \includegraphics[width=5cm]{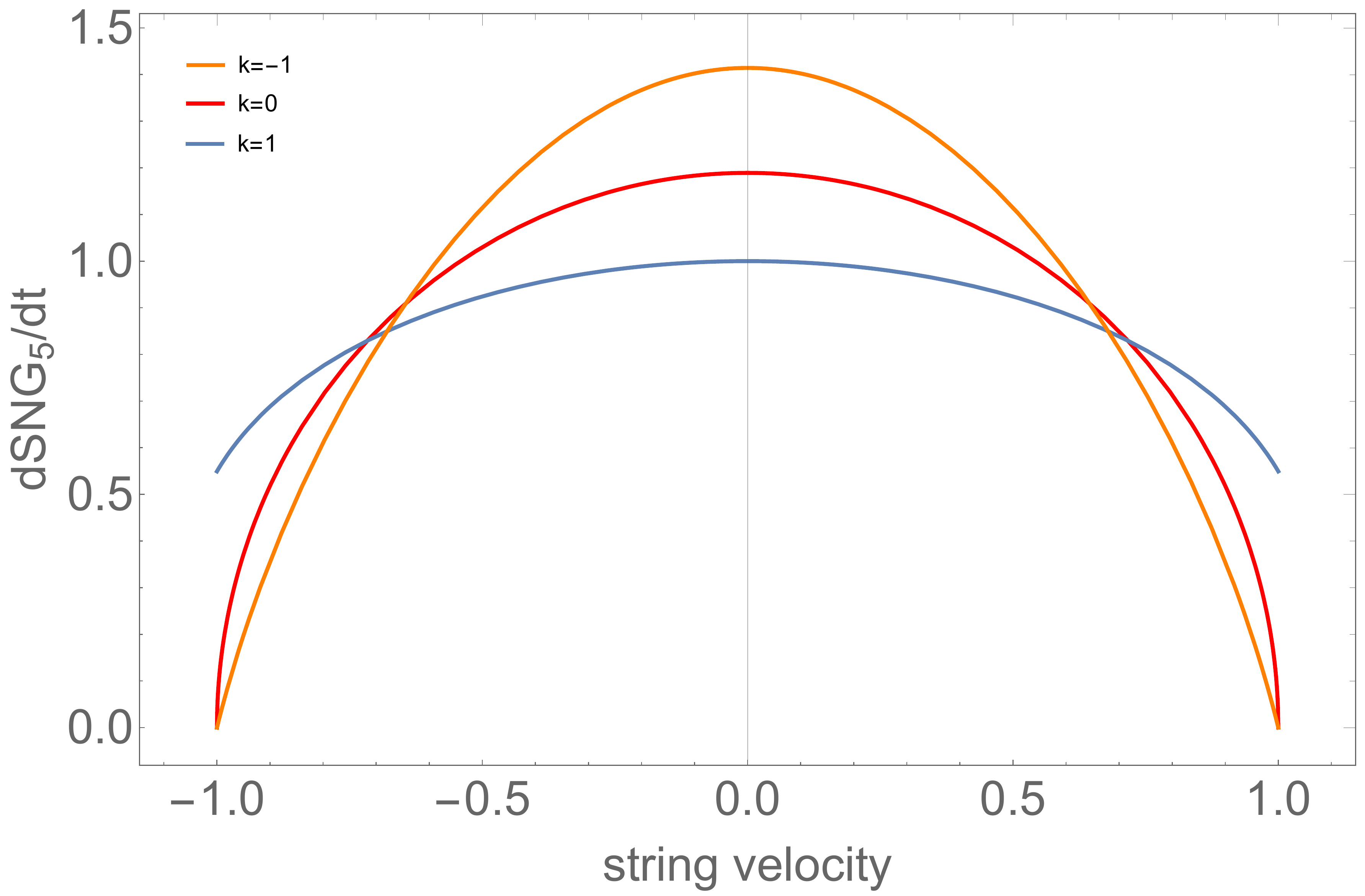}\ \hspace{0.1cm}
   \includegraphics[width=5cm]{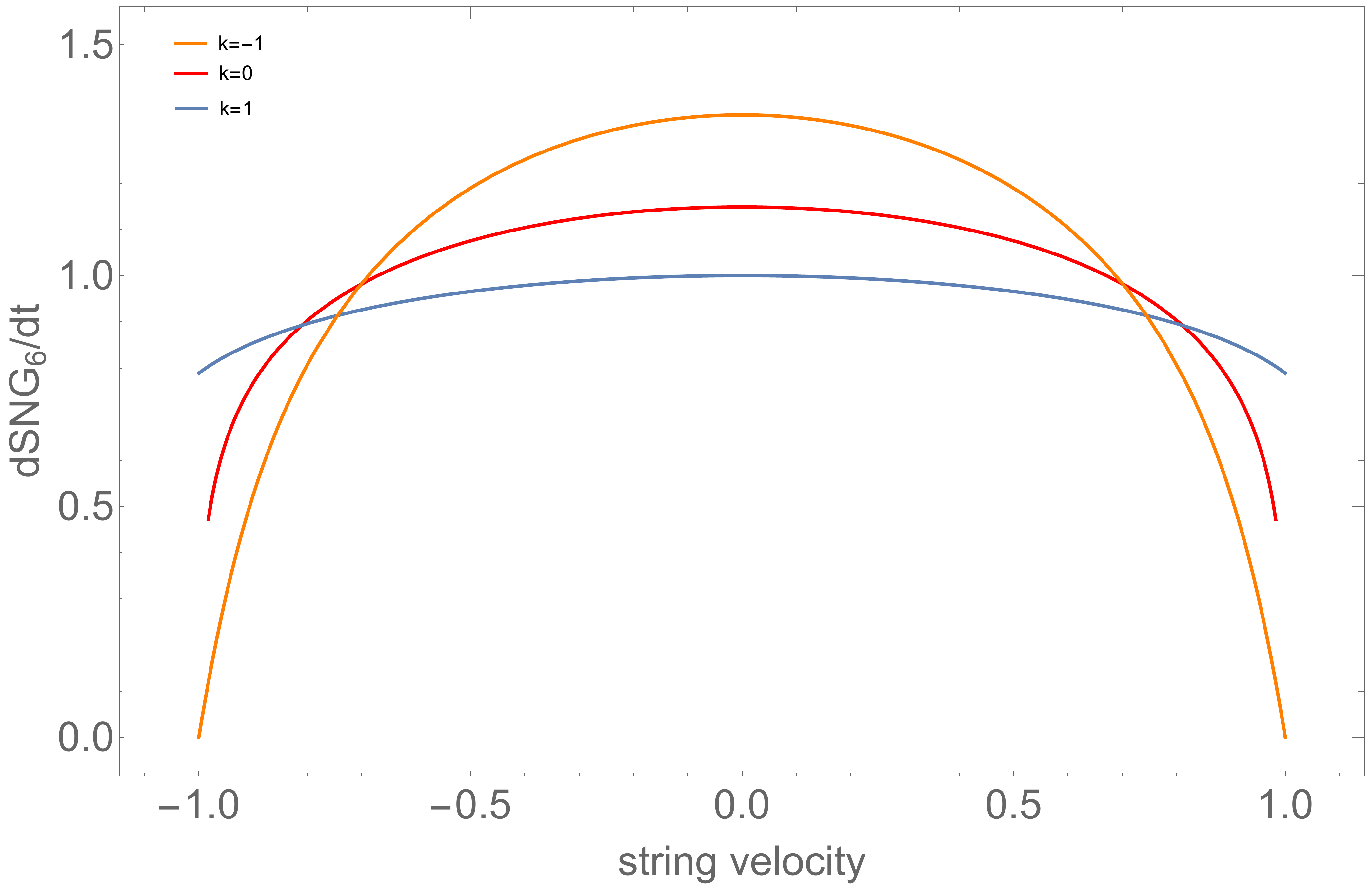}
    \includegraphics[width=5cm]{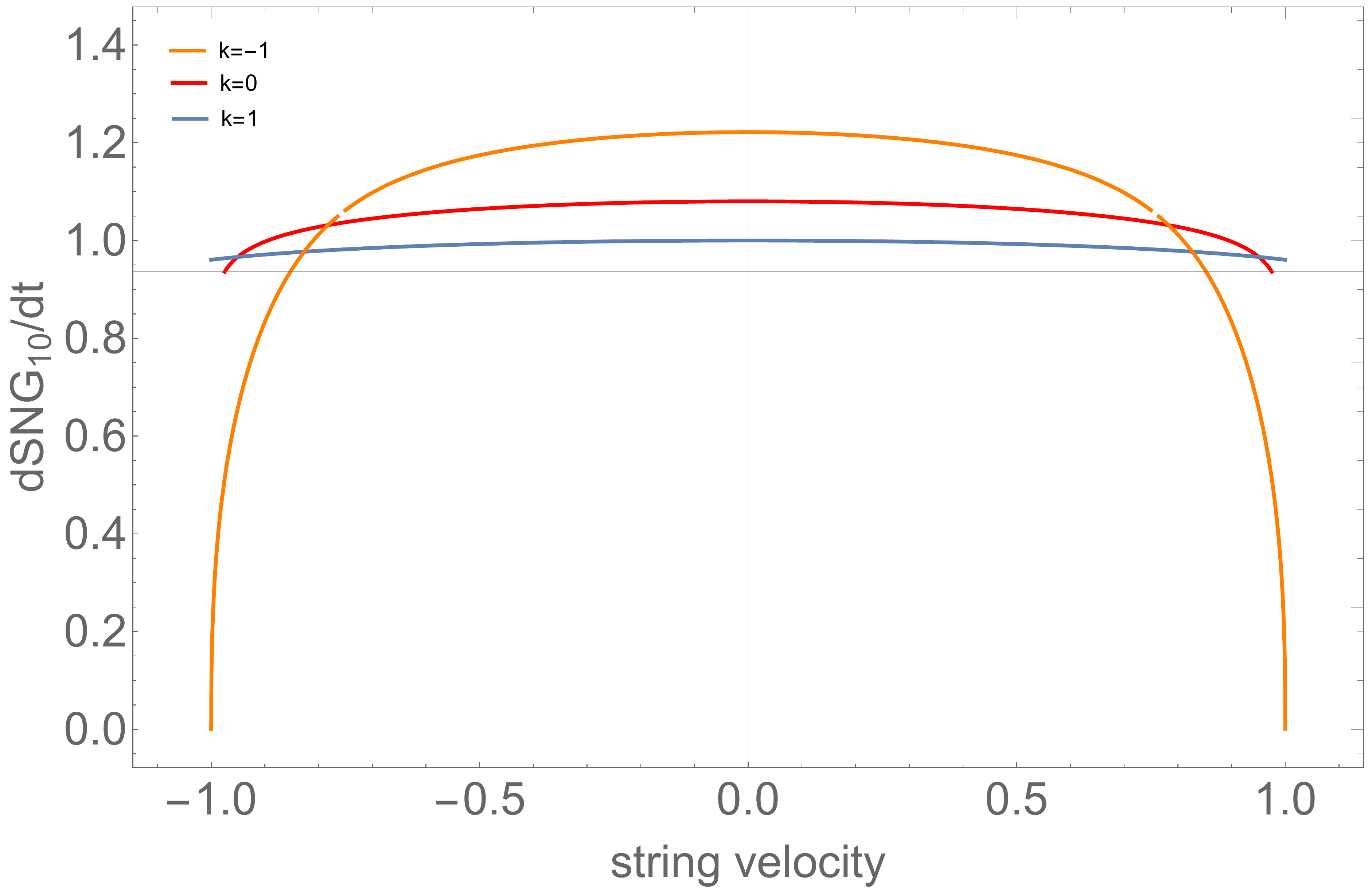}
	  \caption{{Action growth v.s. String velocity} for AdS black hole. From left to right, the dimension is $5,6$ and $10$. In all cases, we have set the black hole mass $m_{0}=2$.}
 \label{fig:dnm0kv}
\end{figure}
%%%%%%%%%%%
%%%%%%%%%%%
\begin{figure}[ht!]
 \centering
  \includegraphics[width=5cm]{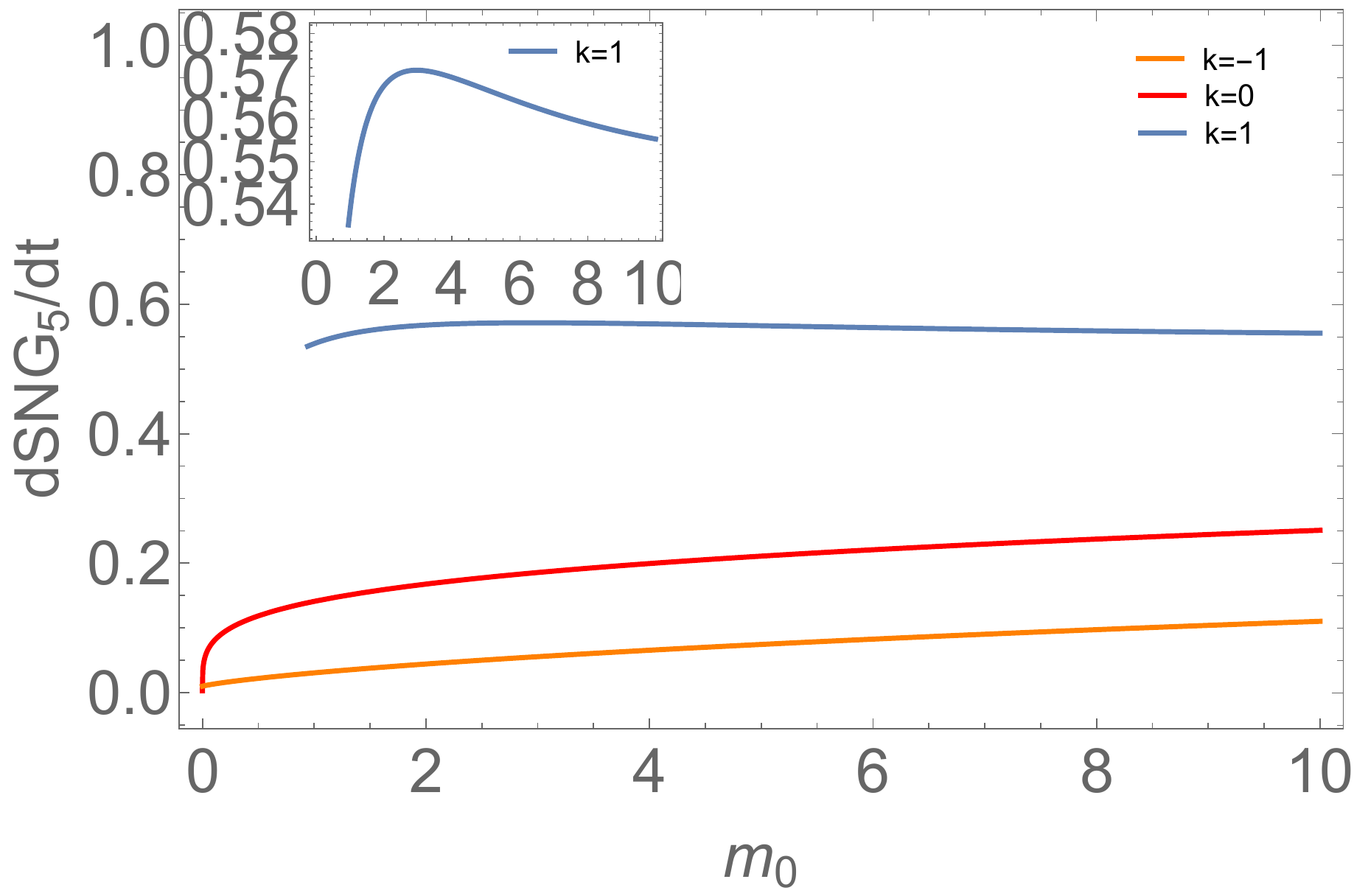}\ \hspace{0.1cm}
   \includegraphics[width=5cm]{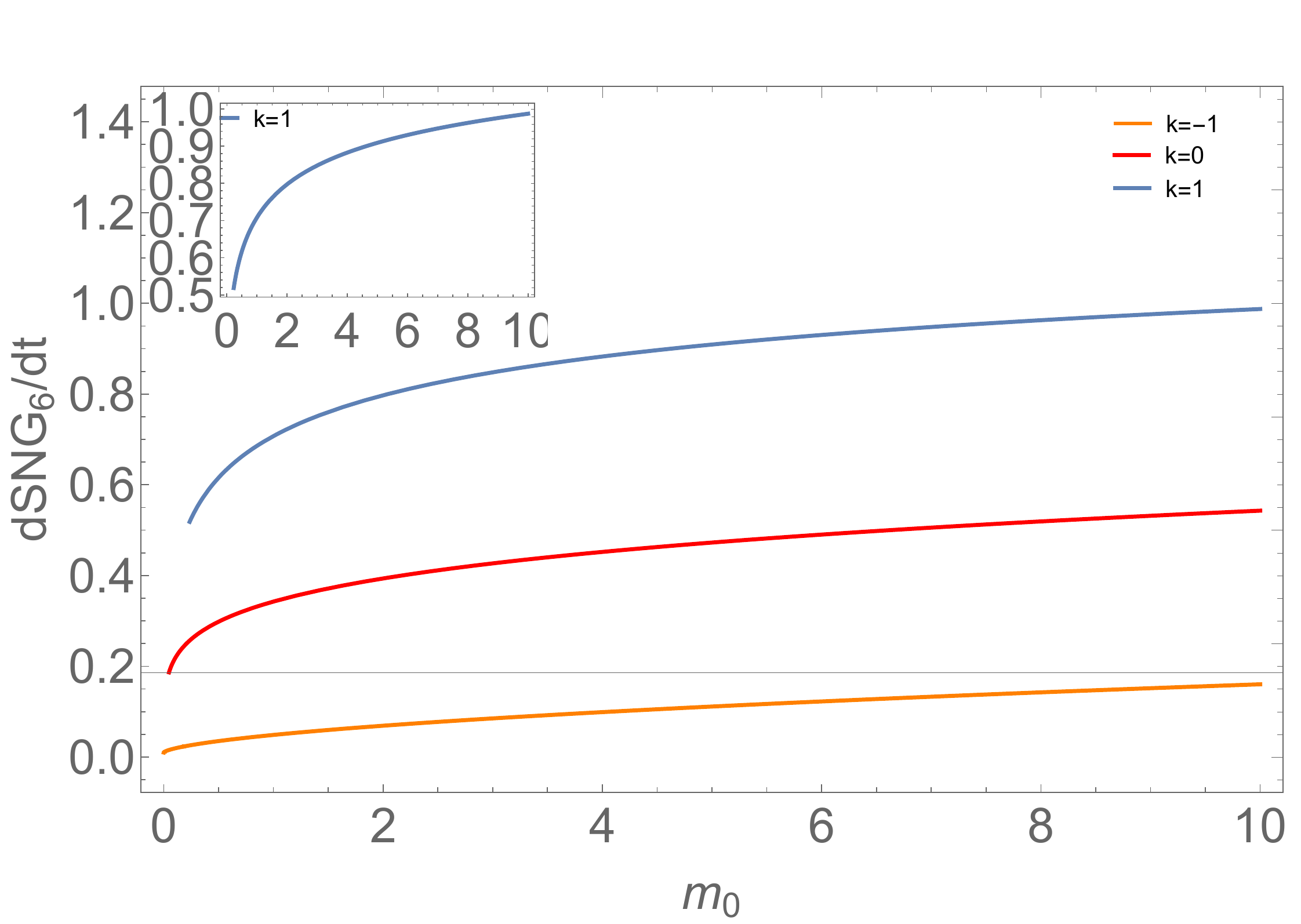}
    \includegraphics[width=5cm]{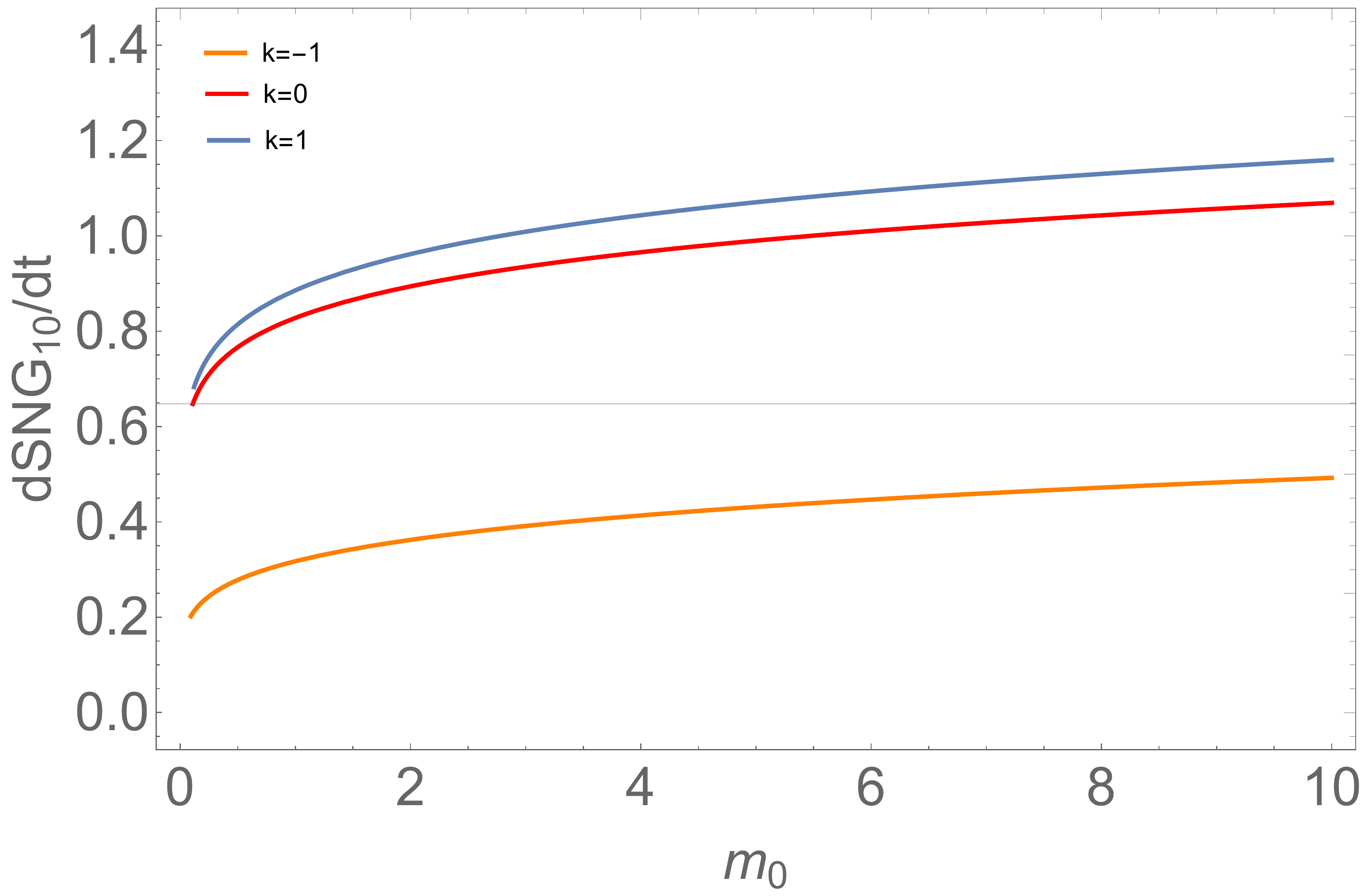}
	  \caption{{Action growth v.s. Black hole mass} From left to right, the dimension is $5,6$ and $10$. In all cases, we have set the $m=0,v=0.99$.}
 \label{fig:dnkm0}
\end{figure}
%%%%%%%%%%%
%%%%%%%%%%%

The velocity dependence and black hole mass dependence in four dimensional case are shown in Fig.\ref{fig:d4m0kv}. In the left plot  with the black hole mass $m_0=2$, the action growth for $k=1$ reproduces the curve with black hole mass $M=1$ in figure 1 of \cite{Nagasaki:2018csh}. In all cases, the maximal value appears when the string does not move, and as the velocity increases, the action growth decreases. It is obvious that the maximal values in the case with $k=-1$ is the largest while it is larger in $k=0$ than in $k=1$ case. This phenomena can be understood in an analytical way. In the limit $v\to 0$, the integration \eqref{NG2} with four dimension $(n=2)$ is
\begin{equation}\label{m0v0actgroth}
\frac{1}{T_{s}}\frac{dS_{NG}}{dt}= r_{h}.
\end{equation}
Then recalling the event horizon satisfying $f(r_{h})=0$, which gives us
\begin{equation}\label{generalkecpress}
k=\frac{m_{0}}{r_{h}}-r^{2}_{h}-m^{2}\left(\frac{r_{h}}{2}+1\right).
\end{equation}
It is obvious that for AdS case with $m=0$ and $m_0=2$, $k$ decreases as $r_h$ increases, which indicates that $k=-1$ corresponds to largest $r_h$, so does the largest action growth while $k=1$ corresponds to the smallest action growth. It is easy to see that the conclusion is also hold in massive gravity with $m\neq0$, which will match the numerical results as we will see soon. Noted that similar analysis could also be done in higher dimension cases.

But the action growth decays fastest when $k=-1$. As the string velocity increases, there exists a {intersection} and the effect of $k$ becomes opposite to that in small velocity. Another notable feature is that for $k=0$ and $k=-1$, the action growth with light speed goes to zero, namely the contribution from NG action disappears such that the effect of string is zero in these cases.  This phenomena is in contrast to the nonzero value for $k=1$. As we increase the dimension(see Fig.\ref{fig:dnm0kv} for 5D,6D and 10D cases.), the action growth from left to right becomes gentler and gentler as usual, and the {intersection} occurs at larger velocity. Moreover, similar as $k=1$, the action growth near light speed for $k=0$ also increases to be nonzero for higher dimensions, while for $k=-1$, it keeps zero.

In the right plot of Fig.\ref{fig:d4m0kv}, we show how the action growth depends on the black hole mass near the light speed. Again the blue curve for $k=1$
reproduces the curve with the same velocity in  figure 1 of \cite{Nagasaki:2018csh}, and there exists a peak at small black hole mass. In a contrast,
it behaves as a monotonically increasing function of the black hole mass for $k=-1$ and $k=0$. Moreover, it shows that in the vicinity of light speed, the action growth for $k=1$ is the largest, which is consistent with that in the left plot.

As the dimension increases (see Fig.\ref{fig:dnkm0} for $5D,6D$ and $10D$ cases), the peak for $k=1$ could  disappear  and the black hole mass dependence becomes monotonically increasing function, which reproduces the outcome of \cite{Nagasaki:2018csh}. Moreover, it also shows that as the dimension increases, the gap between $k=0$ and $k=-1$ is wider while the gap  between $k=0$ and $k=-1$ is narrower. Unfortunately, it is difficult to find analytical insight of the features in the limit $v\to 1$ even in four dimensional case.

\subsection{Massive AdS black hole}
%%%%%%%%%%%%
%%%%%%%%%%%%
Then, we turn on $m$ to study the effect of the graviton mass and horizon curvature.

\begin{figure}[ht!]
 \centering
  \includegraphics[width=5cm]{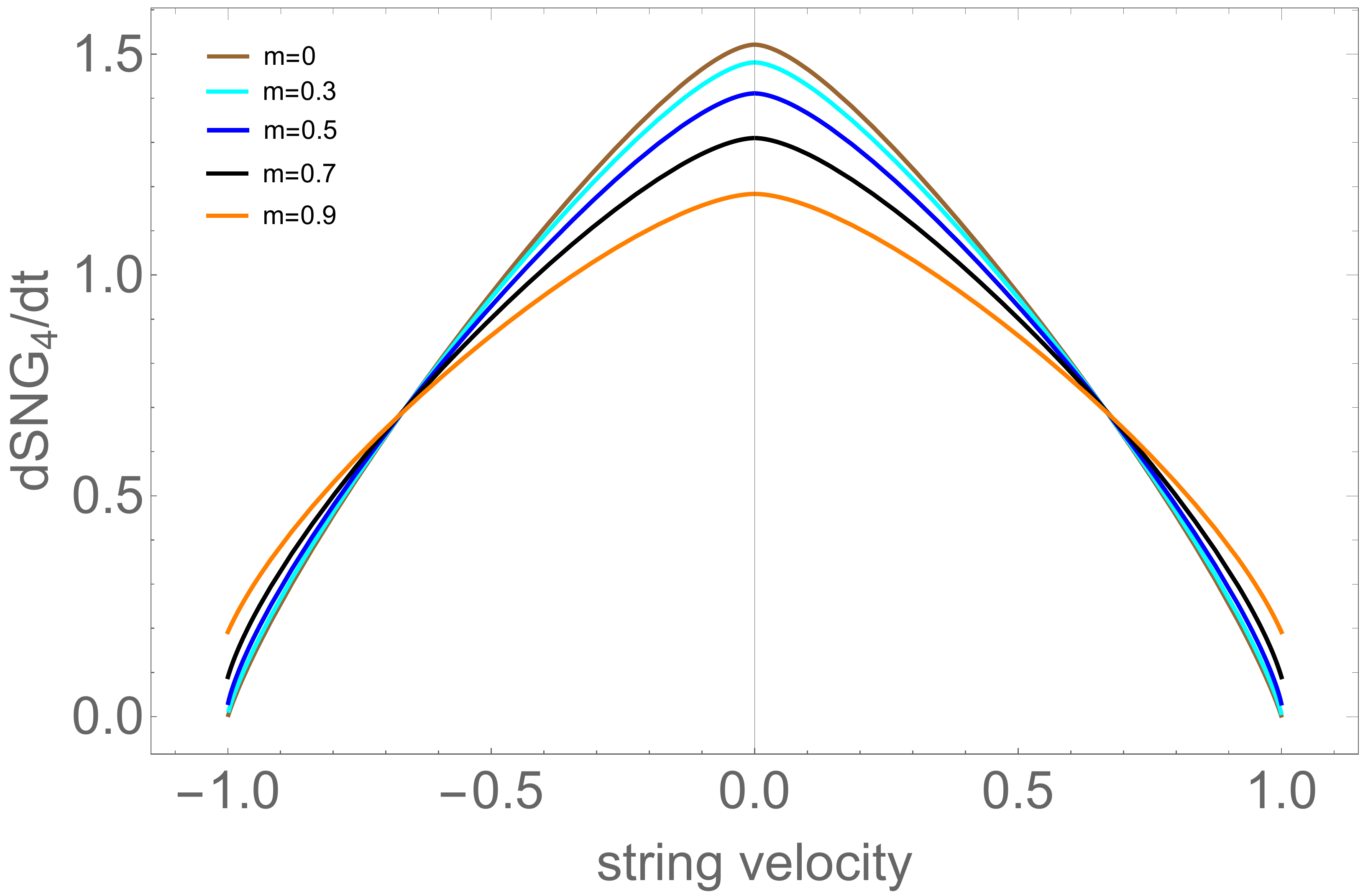}\ \hspace{0.1cm}
  \includegraphics[width=5cm]{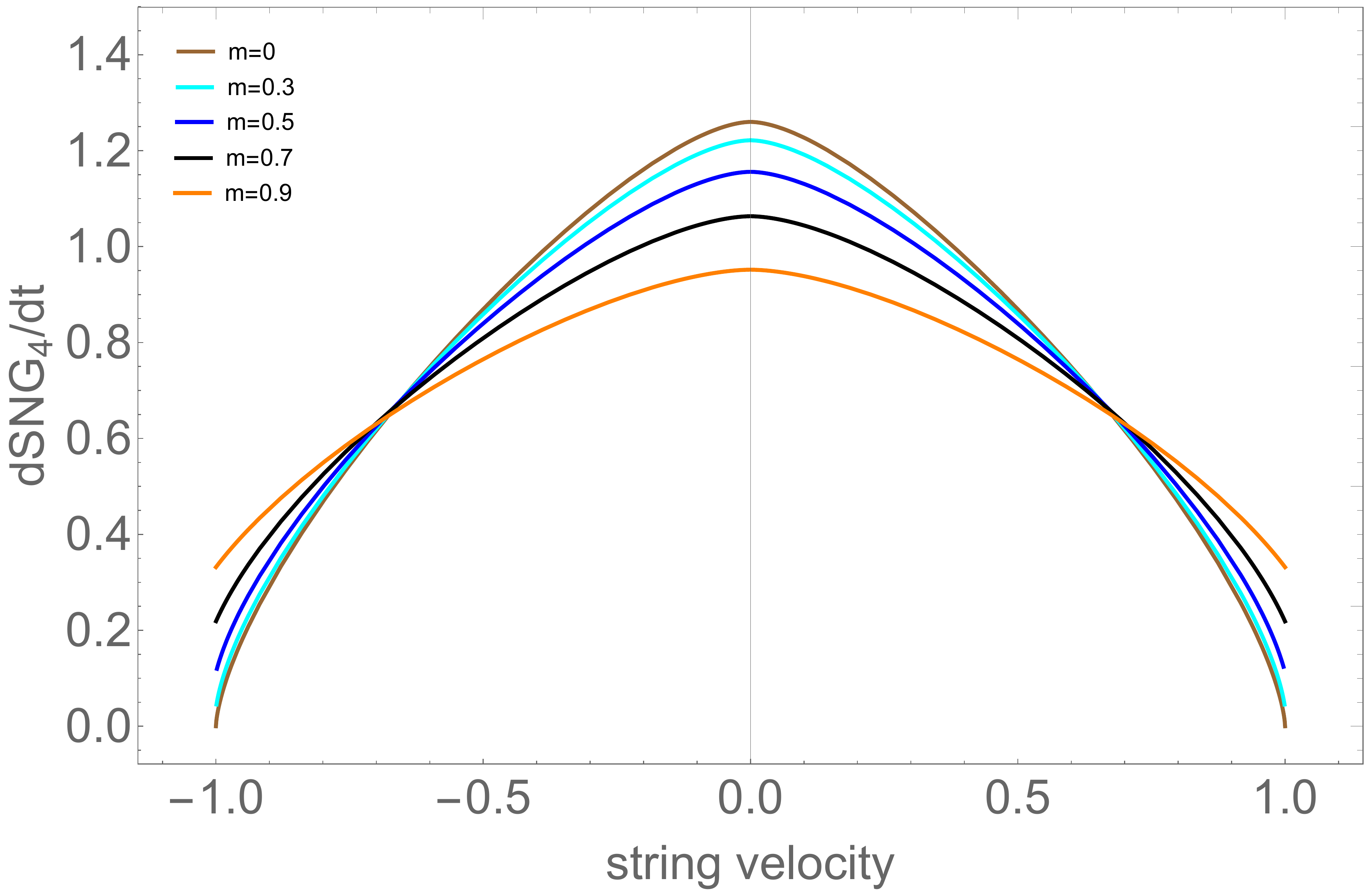}\ \hspace{0.1cm}
  \includegraphics[width=5cm]{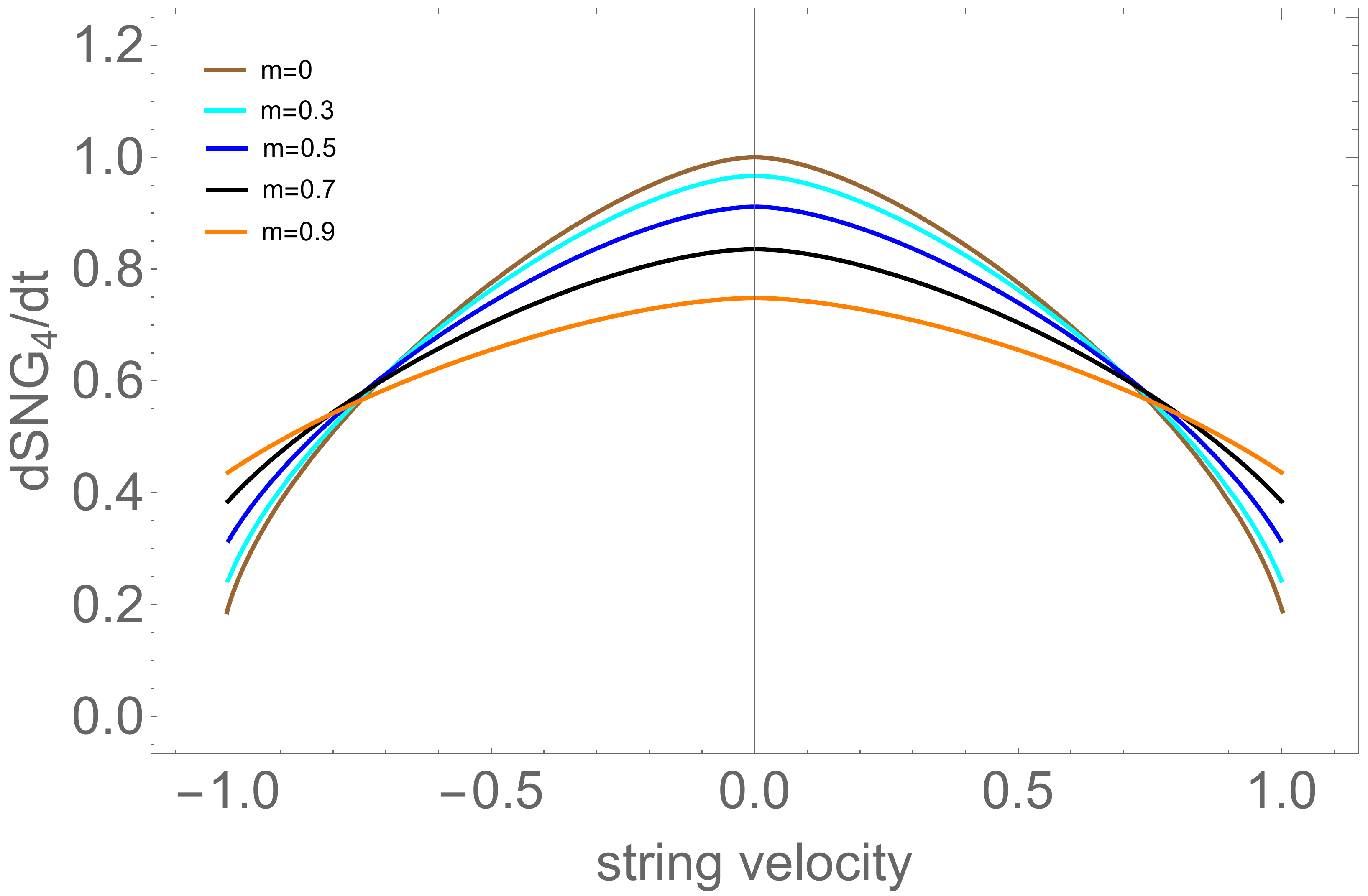}
	  \caption{{Action growth v.s. String velocity} for massive AdS$_4$ black hole with $m_{0}=2$. From left to right, the horizon curvature is $k=-1$, $k=0$ and $k=1$, respectively.}
 \label{fig:d4mv}
\end{figure}

\begin{figure}[ht!]
 \centering
  \includegraphics[width=5cm]{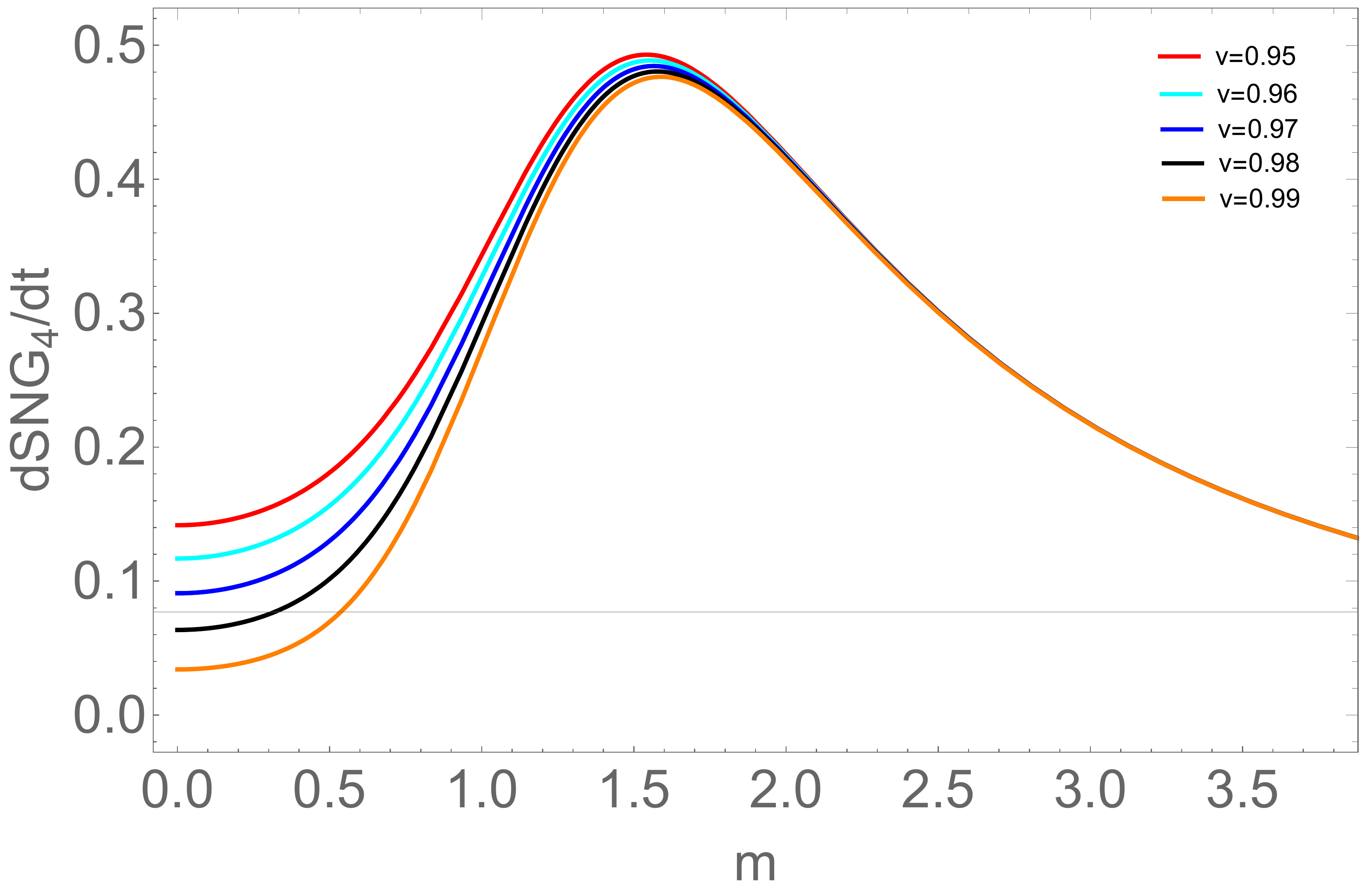}\ \hspace{0.1cm}
  \includegraphics[width=5cm]{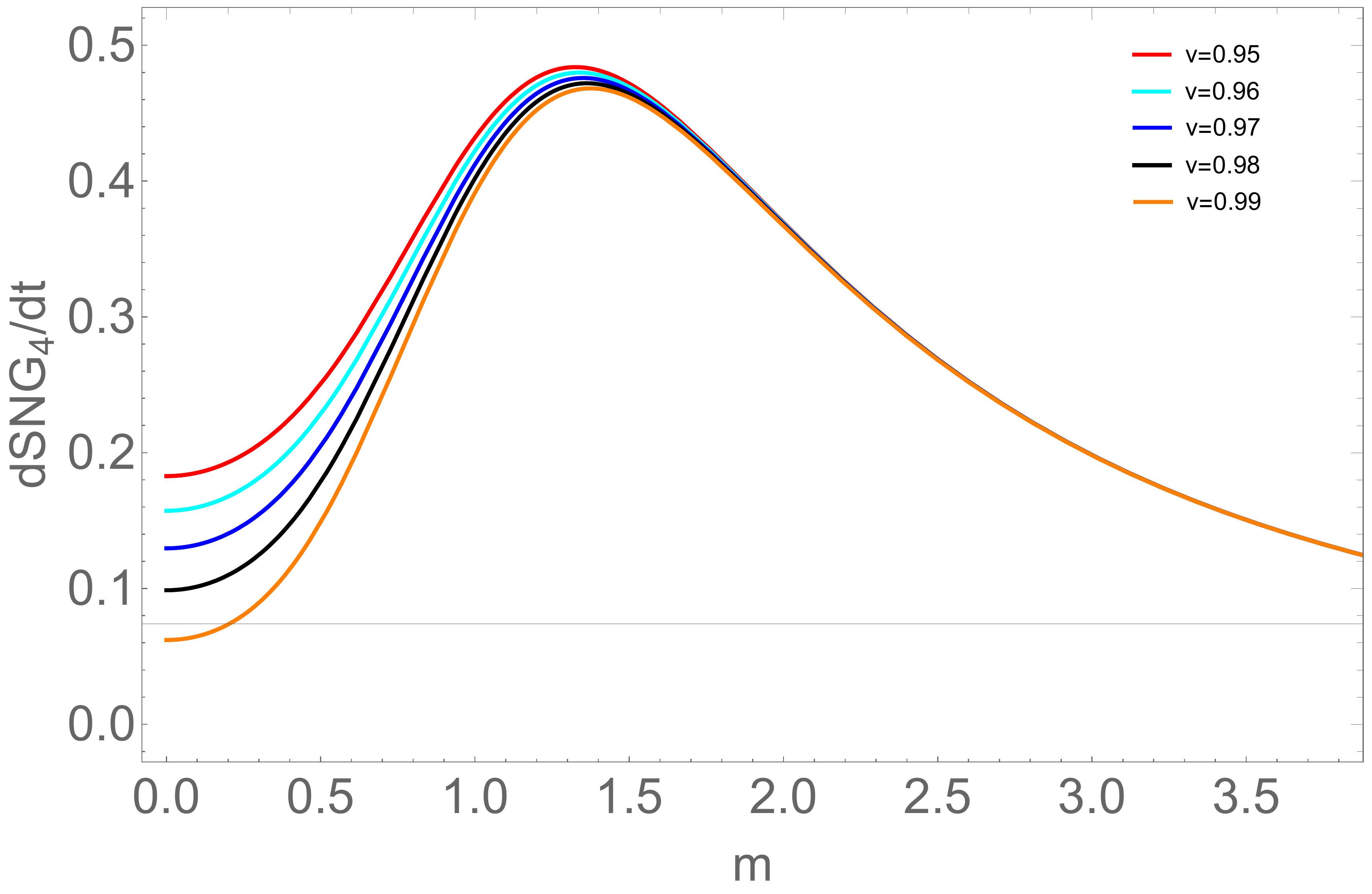}\ \hspace{0.1cm}
  \includegraphics[width=5cm]{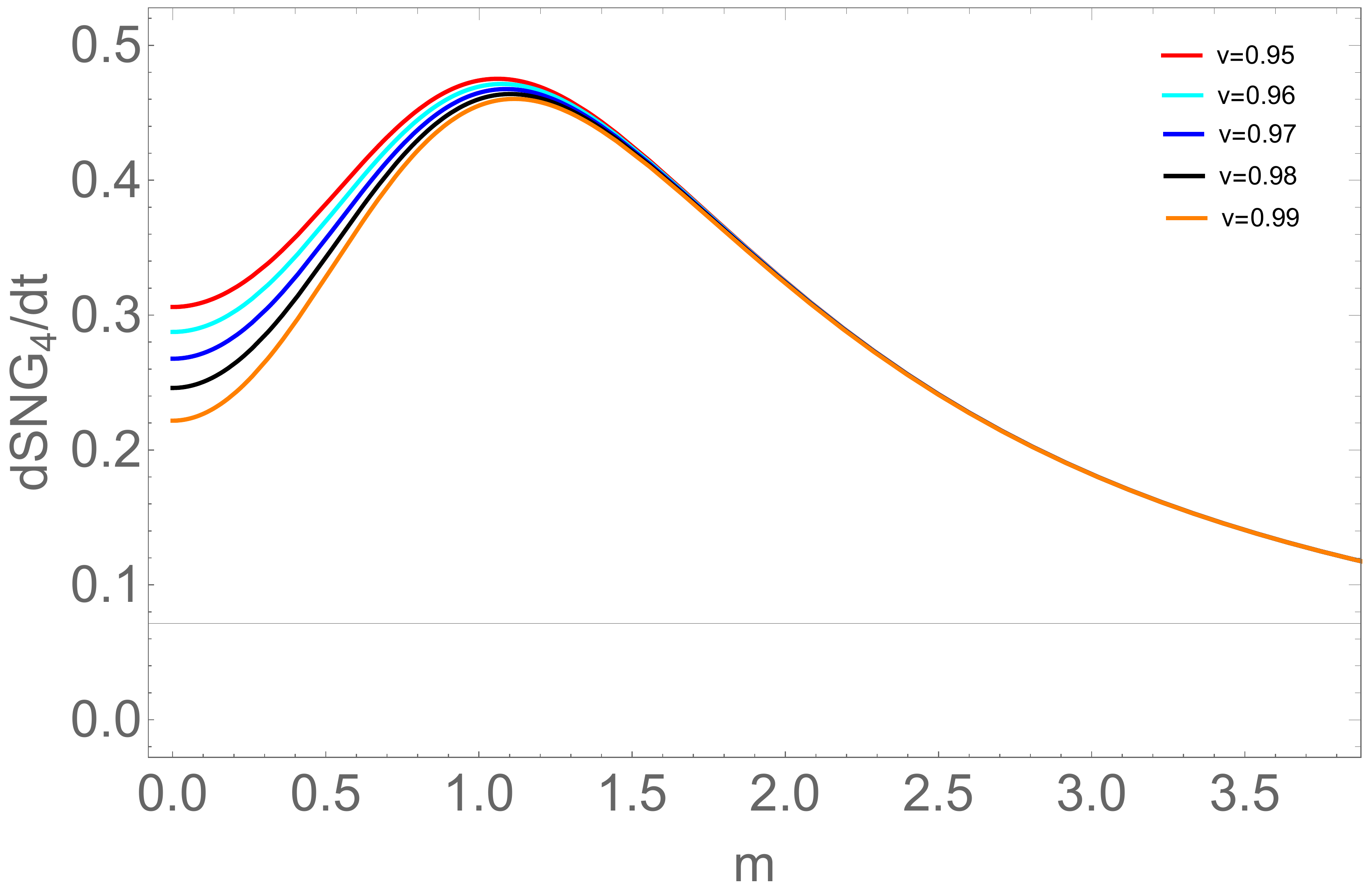}
	  \caption{{Action growth v.s. Graviton mass} for massive AdS$_4$ black hole with $m_{0}=2$. From left to right, the horizon curvature is $k=-1$, $k=0$ and $k=1$, respectively.}
 \label{fig:d4vm}
\end{figure}

We first study in detail the effect of graviton mass on the string velocity dependence as well as graviton mass dependence in the vicinity of light speed. We show the results for four dimensional case  in Fig.\ref{fig:d4mv} and Fig.\ref{fig:d4vm}, respectively. In all plots of Fig.\ref{fig:d4mv}, larger $m$ gives the lower maximal values of action growth.  This feature could be expected from \eqref{m0v0actgroth} in $v\to 0$ limit if we reform \eqref{generalkecpress} as $m^2=(m_{0}/r_{h}-r^{2}_{h}-k)/(r_{h}/2+1)$, which indicates that with fixed $m_0$ and $k$, larger $m$ corresponds to smaller $r_h$.
There exists a {intersection} where the effect of $m$ on the string velocity dependence will change. This observe is consistent with that in three dimensional case (see Fig.\ref{fig:d3k0}), even though here the curves are more gentler.

Fig.\ref{fig:d4vm} shows the $m$ dependence of action growth in the vicinity of light speed. In all cases, the smaller string velocity corresponds to larger action growth when the graviton mass is small. As the increase of $m$, it approaches to a maximal value and then decreases. The $m$ dependence is slightly affected by the string velocity for large enough $m$. Though the behavior is similar to that for 3D case studied in previous section, we see that the location of peak  tends to smaller $m$.
%%%%%%%%%%
%%%%%%%%%%
%%%%%%%%%%
%%%%%%%%%%
%%%%%%%%%%
%%%%%%%%%%

\begin{figure}[ht!]
 \centering
  \includegraphics[width=7cm]{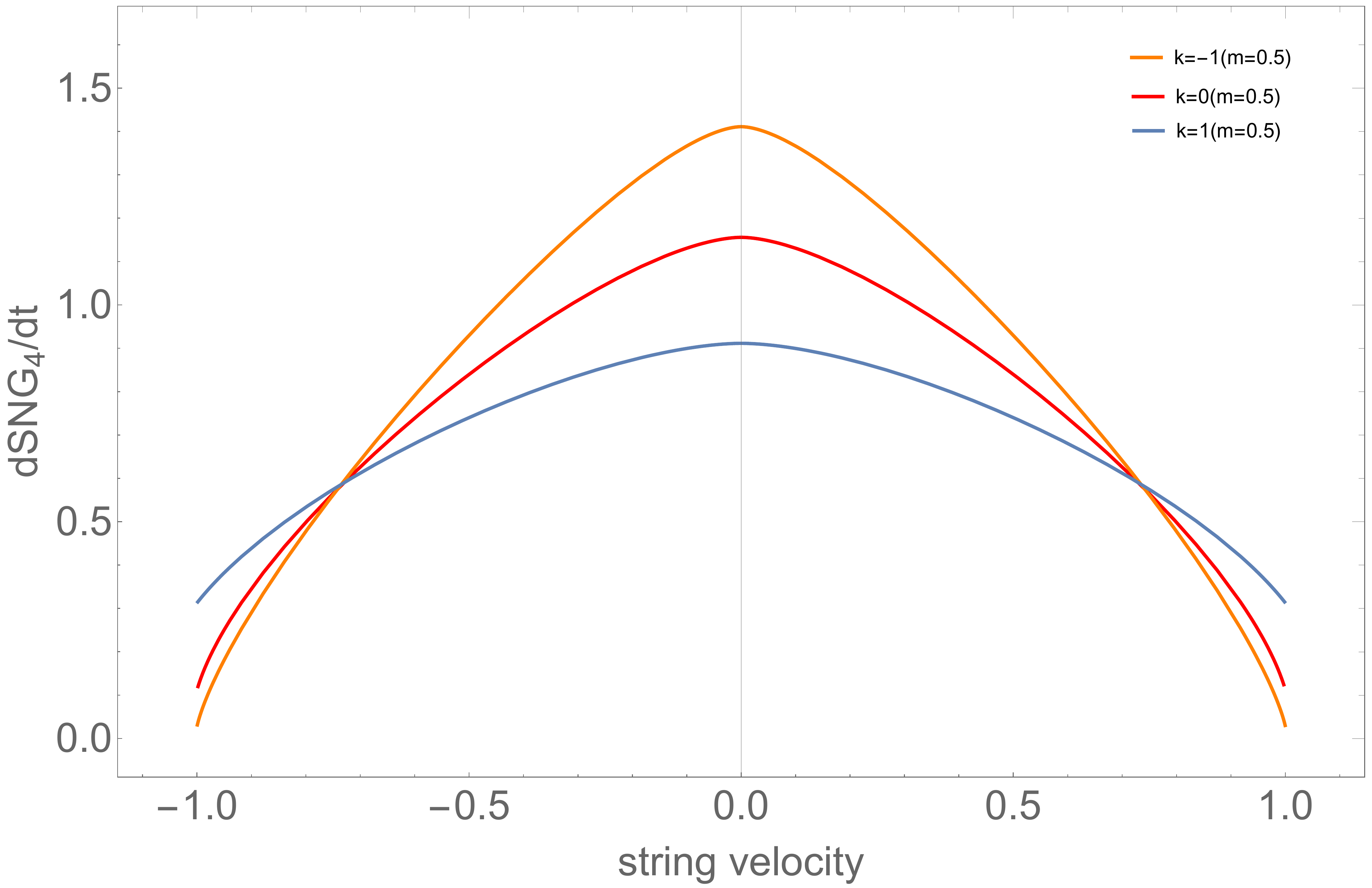}\ \hspace{0.1cm}
   \includegraphics[width=7cm]{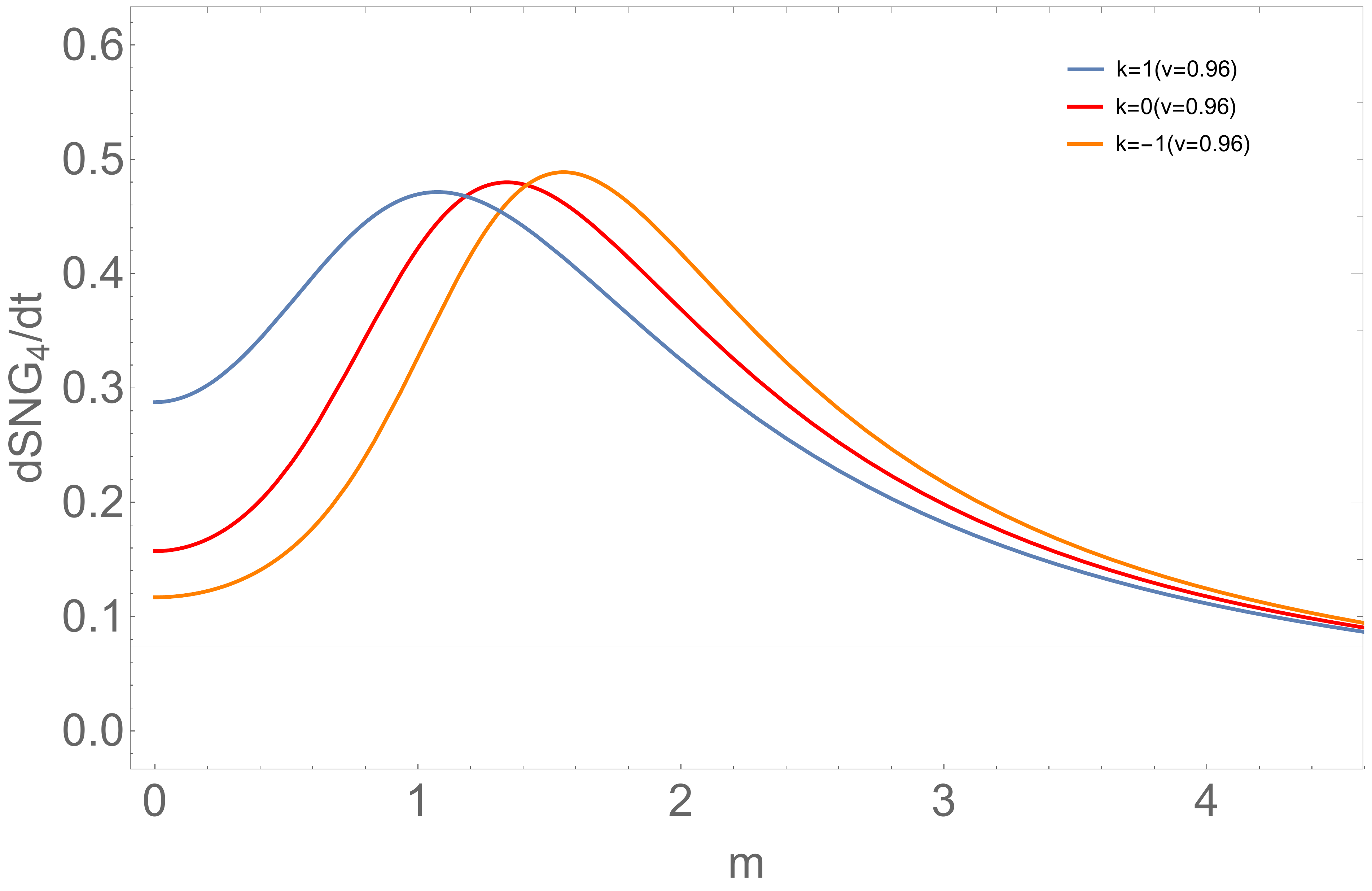}
	  \caption{Action growth for massive AdS$_4$ black hole. Left: {Action growth v.s. String velocity} with $m=0.5$.
	  Right: {Action growth v.s. Graviton mass} with $v=0.96$.
	  Here we have set black hole mass $m_{0}=2$.}
 \label{fig:d4kvm}
\end{figure}

\begin{figure}[ht!]
 \centering
  \includegraphics[width=5cm]{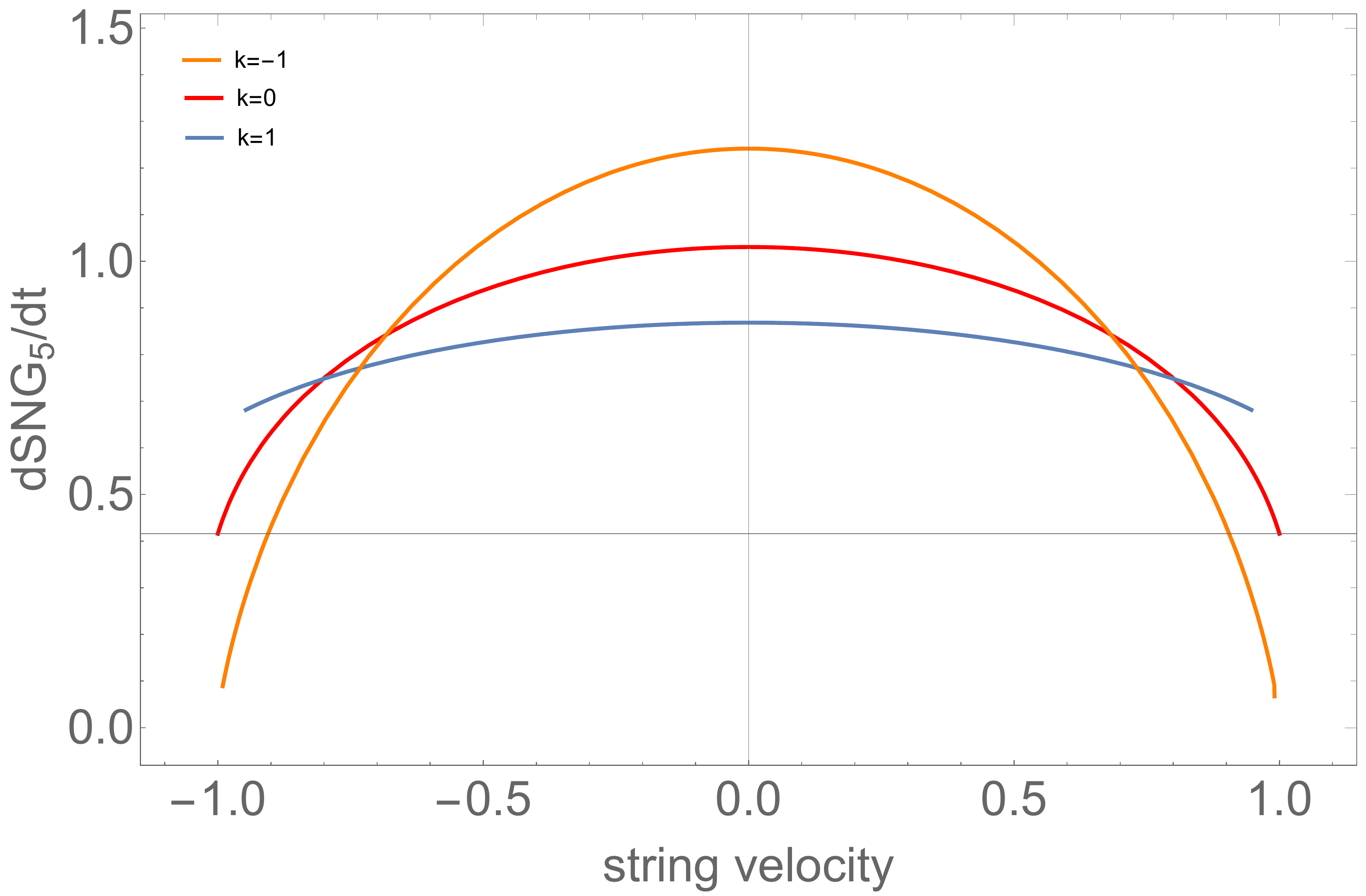}\ \hspace{0.1cm}
   \includegraphics[width=5cm]{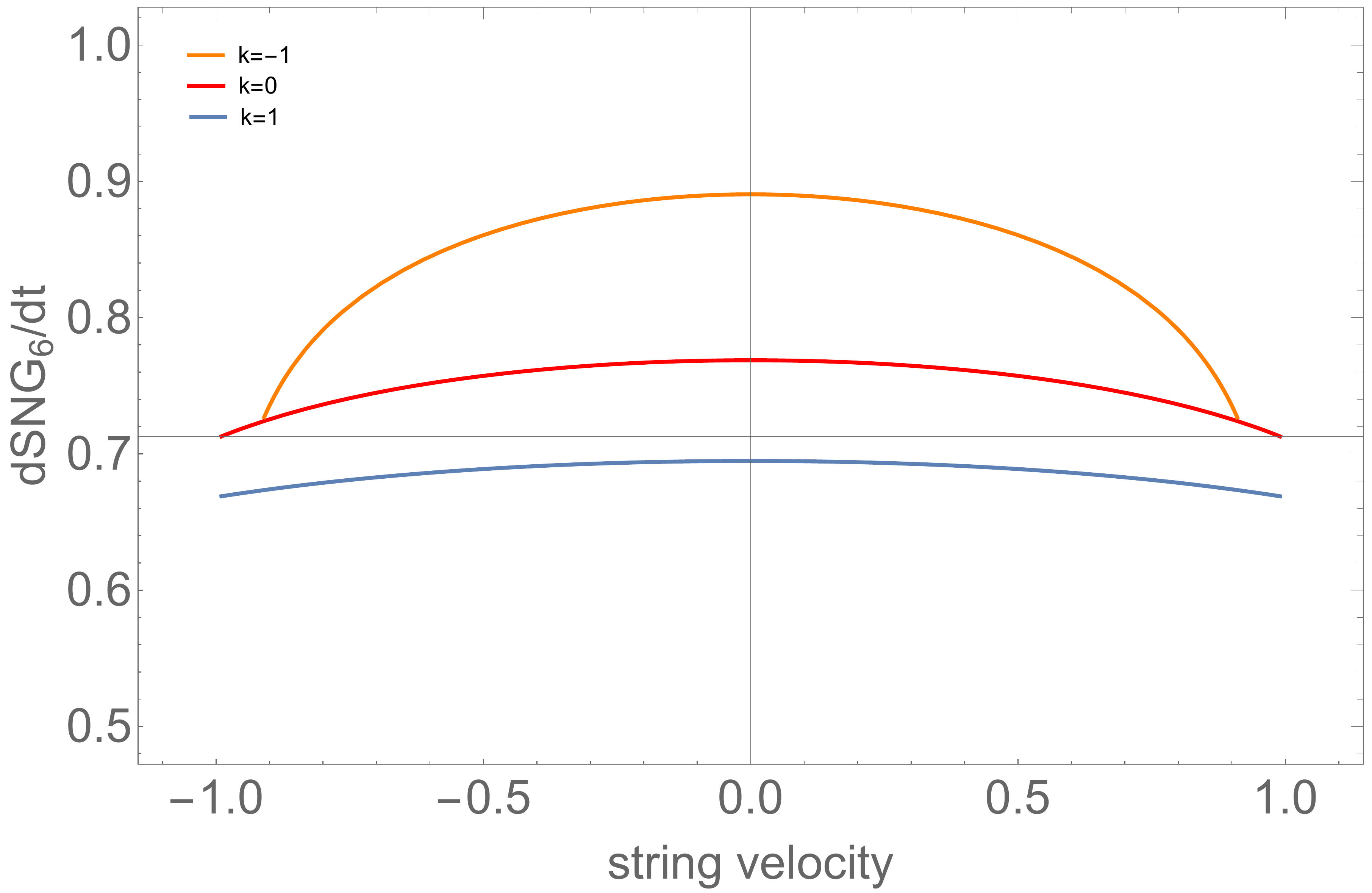}
   \includegraphics[width=5cm]{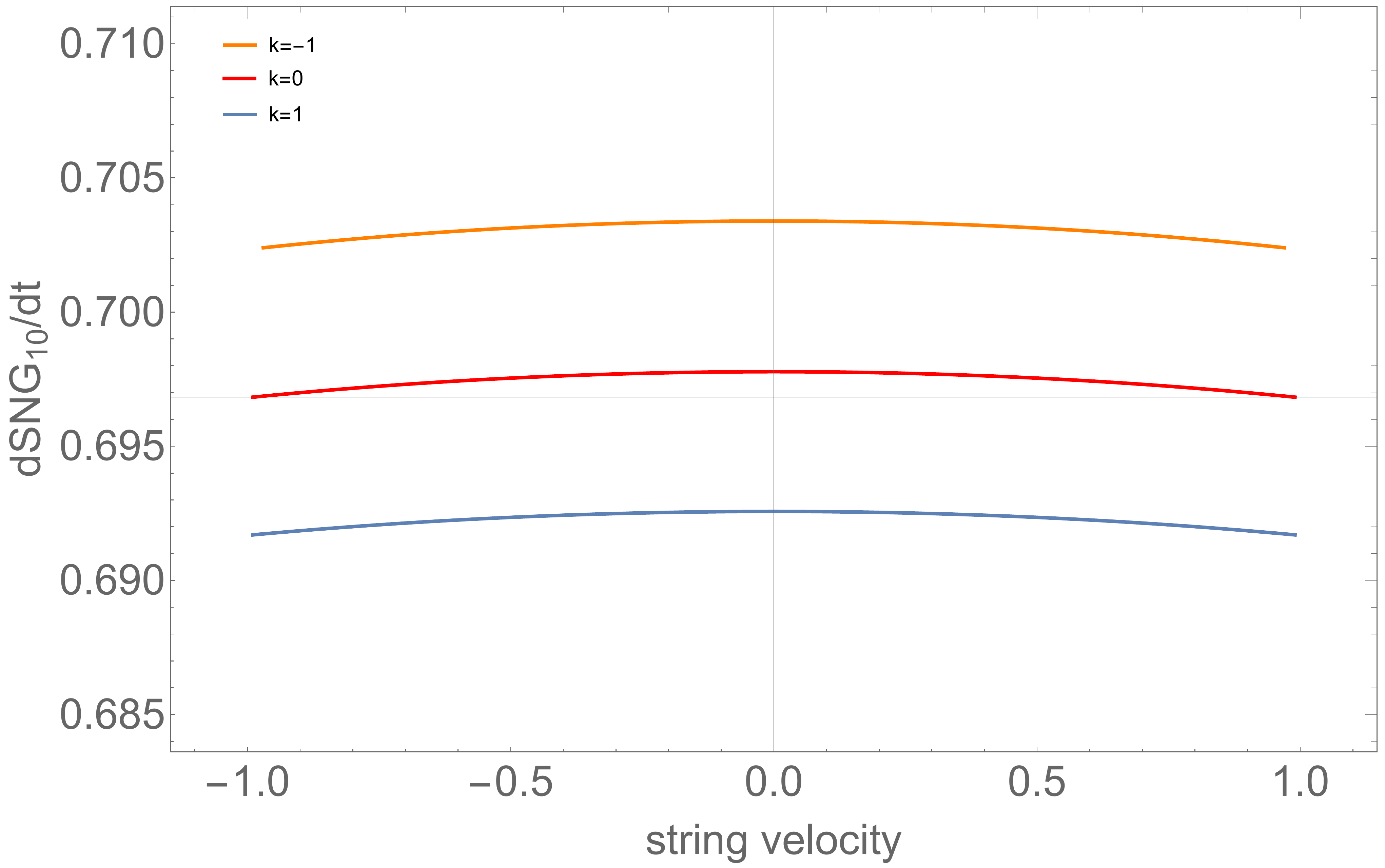}
	  \caption{{Action growth v.s. String velocity} in massive AdS black hole with $m=0.5$.
	  From left to right, the dimension is $5,6$ and $10$.
	  In all cases, we have set the black hole mass $m_{0}=2$.}
 \label{fig:dnkv}
\end{figure}

Then we fix $m=0.5$ and study the effect of horizon curvature. The results in four dimensional massive gravity  are shown in Fig.\ref{fig:d4kvm}. Comparing the left plots in Fig.\ref{fig:d4kvm} and Fig.\ref{fig:d4m0kv}, we see that in massive gravity,
the effect of horizon curvature on the velocity dependent action growth is similar to that in Einstein gravity.
As in the AdS case, the maximal value for stationary string with $k=-1$ is the largest, the reason of which has been stated below equation \eqref{generalkecpress}. There also exists a {intersection.} As the dimension increases (see Fig.\ref{fig:dnkv}), the velocity dependent action growth is more gentler. One new feature in massive gravity is that in all cases, the action growth in light speed limit becomes non-vanished. Comparing with that in AdS case, the {intersection} in massive case for different curvatures could separate in lower dimension.
%%%%%%%%
%%%%%%%%
%%%%%%%%%%%%
%%%%%%%%%%%%
%%%%%%%%%%%%
%%%%%%%%%%%
%%%%%%%%%%%

\begin{figure}[ht!]
 \centering
  \includegraphics[width=5cm]{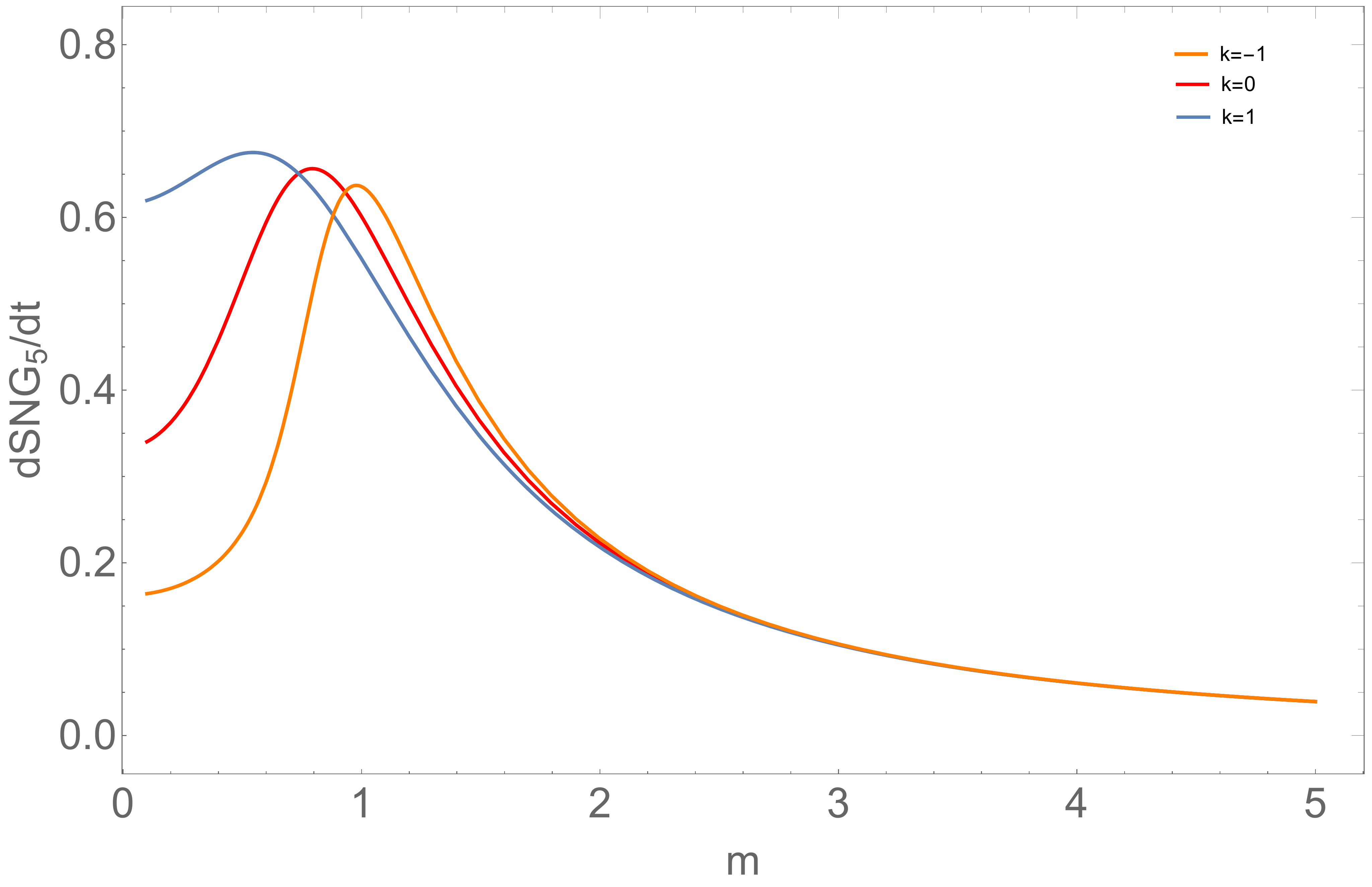}\ \hspace{0.1cm}
    \includegraphics[width=5cm]{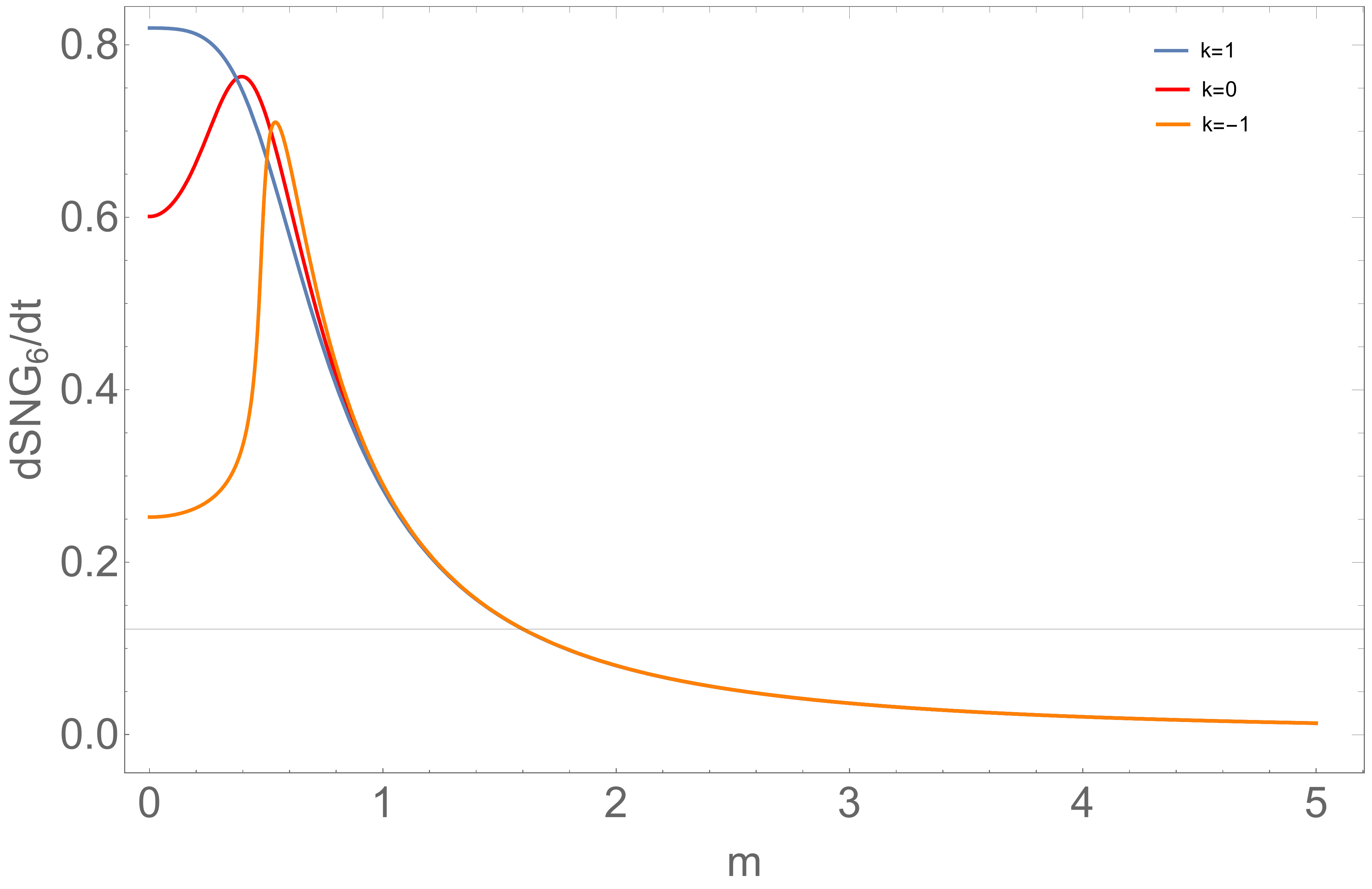}
   \includegraphics[width=5cm]{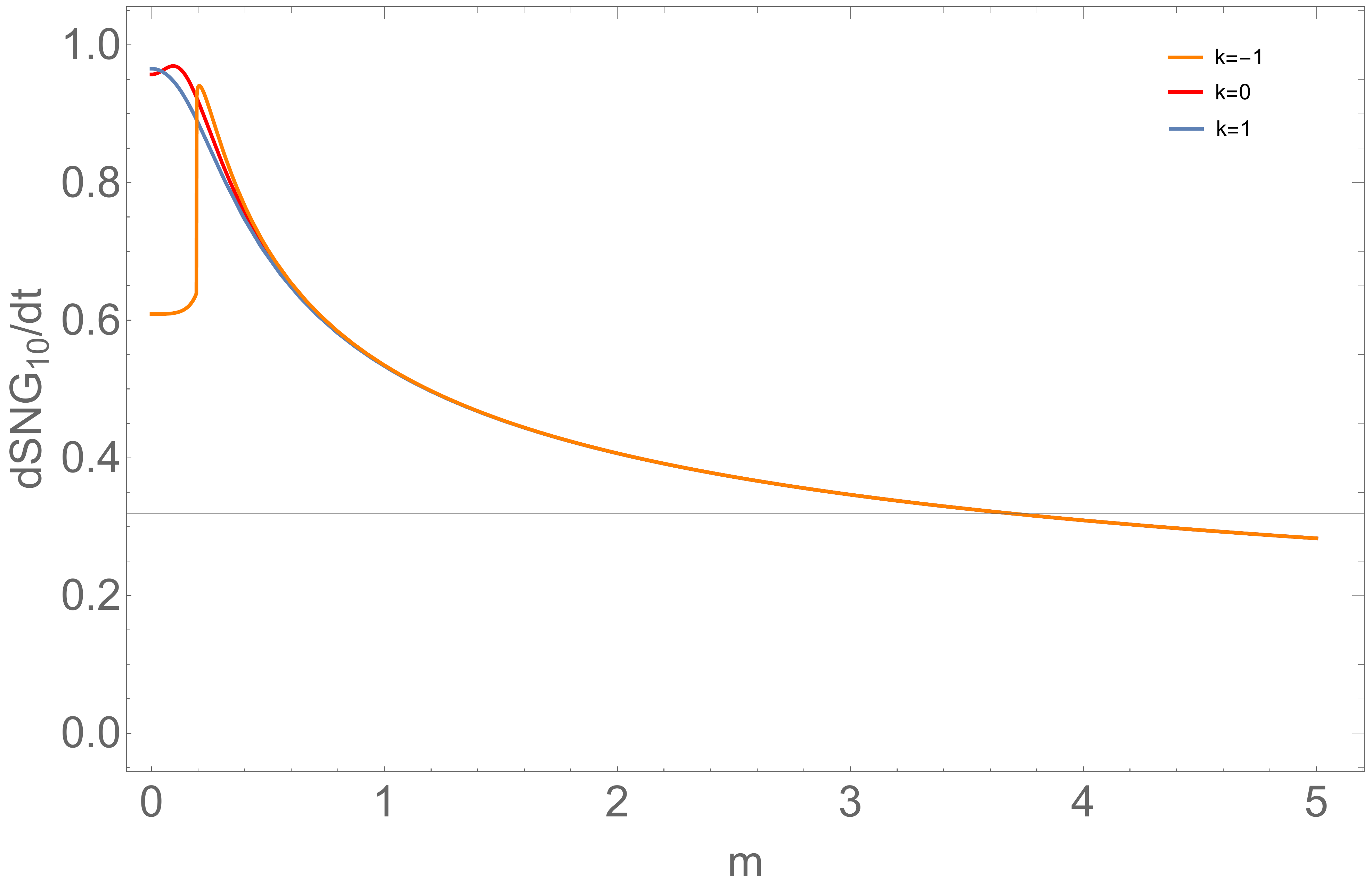}
	  \caption{{Action growth v.s. Graviton mass} for massive AdS black hole
	  From left to right, the dimension is $5,6$ and $10$.
	  In all cases, we have set the black hole mass $m_{0}=2$ and the string speed is $v=0.96$.}
 \label{fig:dnkm}
\end{figure}

In the right plot of Fig.\ref{fig:d4kvm}, we show the effect of $k$ on the graviton mass dependence, which is more significant for small $m$ than for large $m$. It also shows a peak at certain critical $m_c$ as in three dimensional case.  Here the location value of $m$ for the peak decreases as $k$ increases. Furthermore, in Fig.\ref{fig:dnkm} we show the graviton mass dependence for higher dimensions.
As the dimension increases, $m$ for the location of peak decreases. Then for $k=1$, the peak shows at $m=0$, meaning the action grow monotonically decreases as $m$ increases(see the middle plot for six dimension), which is also explicit in the ten dimensional case. Further increasing the dimension, the action grow could also become monotonically decreasing function of $m$ for the massive black hole with $k=-1$ and $k=0$. This is a new phenomena.
%%%%%%%%%%%
%%%%%%%%%%%
%%%%%%%%
%%%%%%%%
\section{Summary} \label{summ}
We studied the complexity growth in the dynamical system with a Wilson line operator, which is holographically dual to the massive black holes with a probe string. We focused on the Nambu-Goto action growth and  employed the CA conjecture to explore the effect of string on the complexity growth in a boundary gauge theory with momentum relaxation. We mainly analyzed  the effect of horizon curvatures, the graviton mass and the dimension on the velocity dependent complexity growth. Our study is mainly presented in numerical way accompanying with some analytic estimation. The results are summarized as follows.

In three dimension, the maximal complexity growth appears when the string is motionless, and  as the graviton mass increases, the maximal value becomes smaller. As the string moves faster, the action growth becomes smaller. When the string moves with light speed, the action growth in massive BTZ black hole will not vanish which is different from that in BTZ black hole.
In the vicinity of the light speed, as the graviton mass increases, the action growth first increases and then decreases. Moreover, the velocity dependence is significant for small graviton mass, while it is slight for large $m$. It means that in the vicinity of light speed, the graviton mass suppresses the velocity dependence of the action growth, which is very different from the effect of BTZ black hole mass that enhances it.

In higher dimensions, we could study the effect of horizon curvatures $k$. We first studied  the effect of $k$ on the velocity dependent action growth. Our results shows that the maximal value
for the stationary string for $k=-1$ is the largest, then for $k=0$, the value for $k=1$ is the smallest. There is a {intersection} in the velocity dependence curves for different $k$. As the dimension increases, the velocity dependent action growth is more gentler.
Comparing with AdS black hole, in massive black hole, the action growth in light speed limit becomes non-vanishing with different $k$, and the {intersection} for different curvatures could separate in lower dimensions.

We then studied the effect of $k$ on the graviton mass dependence. It is more significant for small $m$ than for large $m$. It also shows a peak at certain graviton mass as in three dimensional case, and the location value of the peak decreases as $k$ increases.
As the dimension increases, the critical graviton mass for the peak decreases.  The peak for $k=1$ in high enough dimension could first appears at $m=0$, meaning the action growth turns to monotonically decrease as $m$ increases. Then further increasing the dimension, the action growth could also become monotonically decreasing function of $m$ for the massive black hole with $k=-1$ and $k=0$.

Inspired by Lloyd's state on the quantum complexity
growth rate\cite{Lloyd}, the authors of \cite{Brown:2015bva} addressed an analogous  ``Lloyd bound"  via CA conjecture as
\begin{equation}\label{Lloyd bound}
\frac{dS}{dt}\leq\frac{2M}{\pi},
\end{equation}
where $S$ is the total action and $M$ is the total mass or energy of the system. In the previous literatures \cite{Pan:2016ecg,Guo:2017rul} where the action consists of Einstein-Hilbert term and boundary term in massive gravity, it was shown that the bound was saturated. In our work with a probe string, the total action consists of Einstein-Hilbert term, boundary term and the NG term, while $M$ contain the contributions of the black hole mass and the mass of string. Then the bound could be satisfied because the length of string is infinity stretching to the infinity of the space. It is noticed that in CA conjecture, the bound \eqref{Lloyd bound} may not be always satisfied, see \cite{Carmi:2017jqz,Kim:2017qrq,Couch:2017yil,Moosa:2017yvt,
Ageev:2018nye,Ageev:2019fxn} as examples for the violated cases.

\begin{acknowledgments}
We appreciate Dmitry S. Ageev and Koichi Nagasaki for helpful corresponding. This work is supported by the Natural Science Foundation
of China (Nos.11705161 and 11775036) and Fok Ying Tung Education Foundation (No.171006).
\end{acknowledgments}

\end{document}